\documentclass[fleqn, usenatbib, epsfig]{mnras}
\usepackage{amssymb}
\usepackage{amsmath}
\usepackage{amstext}
\usepackage{deluxetable}
\usepackage{epsfig}
\usepackage{thumbpdf}
\usepackage{epstopdf}
\usepackage{graphicx}
\usepackage[english]{babel}
\usepackage[T1]{fontenc}
\usepackage[utf8]{inputenc}
\usepackage{ae,aecompl}
\usepackage[lofdepth,lotdepth]{subfig}
\usepackage{natbib}
\usepackage{xargs}                   
\usepackage{txfonts}
\usepackage{multicol}
\usepackage[pdftex,dvipsnames]{xcolor} 
\usepackage[colorinlistoftodos,prependcaption,textsize=tiny]{todonotes}
\usepackage[colorinlistoftodos,prependcaption,textsize=tiny]{todonotes}
\newcommandx{\unsure}[2][1=]{\todo[linecolor=red,backgroundcolor=red!25,bordercolor=red,#1]{#2}} 
\newcommandx{\change}[2][1=]{\todo[linecolor=blue,backgroundcolor=blue!25,bordercolor=blue,#1]{#2}} 
\newcommandx{\info}[2][1=]{\todo[linecolor=OliveGreen,backgroundcolor=OliveGreen!25,bordercolor=OliveGreen,#1]{#2}} 
\newcommandx{\improvement}[2][1=]{\todo[linecolor=Plum,backgroundcolor=Plum!25,bordercolor=Plum,#1]{#2}}
\newcommandx{\thiswillnotshow}[2][1=]{\todo[disable,#1]{#2}}
\usepackage{savesym}

\def\Msun{${\rm M}_{\odot}$}

\DeclareRobustCommand{\ion}[2]{%
\relax\ifmmode
\ifx\testbx\f@series
{\mathbf{#1\,\mathsc{#2}}}\else
{\mathrm{#1\,\mathsc{#2}}}\fi
\else\textup{#1\,{\mdseries\textsc{#2}}}%
\fi}
\def\CO{{\rm CO}}
\def\pasa{{\rm PASA}}
\def\H{H$_2$}

\def\av{\rm A$_V$}
\def\nh{\rm N(H$_2$)}

\title[EDGE SN environments]{Molecular gas at supernova local environments unveiled by EDGE} 

\author[L. Galbany et al.]{L. Galbany$^{1}$\thanks{E-mail: llgalbany@pitt.edu}, 
L. Mora$^{2,3}$, 
S. Gonz\'alez-Gait\'an$^{2,4}$,
A. Bolatto$^{5}$,
H. Dannerbauer$^{6,7}$,
\newauthor
\'A. R. L\'opez-S\'anchez$^{8,9}$,
K. Maeda$^{10,11}$,
S. P\'erez$^{12,3}$,
M. A. P\'erez-Torres$^{13}$,
S. F. S\'anchez$^{14}$,
\newauthor
T. Wong$^{15}$,
C. Badenes$^{1}$
L. Blitz$^{16}$,
R. A. Marino$^{17}$,
D. Utomo$^{16}$,
G. Van de Ven$^{18}$\\
$^{1}$PITT PACC, Department of Physics and Astronomy, University of Pittsburgh, Pittsburgh, PA 15260, USA.\\
$^{2}$Millennium Institute of Astrophysics, Universidad de Chile, Casilla 36-D, Santiago, Chile.\\
$^{3}$Departamento de Astronom\'ia, Universidad de Chile, Casilla 36-D, Santiago, Chile.\\
$^{4}$Centro de Modelamiento Matem\'atico, Universidad de Chile, Av. Blanco Encalada 2120 Piso 7, Santiago, Chile\\
$^{5}$Department of Astronomy, University of Maryland, College Park, MD 20742, USA.\\
$^{6}$Instituto de Astrof\'sica de Canarias (IAC), E-38205 La Laguna, Tenerife, Spain.\\ 
$^{7}$Universidad de La Laguna, Dpto. Astrof\'sica, E-38206 La Laguna, Tenerife, Spain.\\
$^{8}$Australian Astronomical Observatory, P.O. Box 915, North Ryde, NSW 1670, Australia\\
$^{9}$Department of Physics and Astronomy, Macquarie University, NSW 2109, Australia\\
$^{10}$Department of Astronomy, Kyoto University, Kitashirakawa-Oiwake-cho, Sakyo-ku, Kyoto 606-8502, Japan.\\
$^{11}$Kavli Institute for the Physics and Mathematics of the Universe (WPI), The University of Tokyo, 5-1-5 Kashiwanoha, Kashiwa, Chiba 277-8583, Japan.\\
$^{12}$Millenium Nucleus "Protoplanetary Disks in ALMA Early Science", Universidad de Chile, Casilla 36-D Santiago, Chile\\
$^{13}$Instituto de Astrof\'isica de Andaluc\'ia - CSIC, PO Box 3004, 18008 Granada, Spain\\
$^{14}$Instituto de Astronom\'ia, Universidad Nacional Aut\'onoma de M\'exico, A.P. 70-264, 04510 M\'exico, D.F., Mexico.\\
$^{15}$Department of Astronomy, University of Illinois, Urbana, IL 61801, USA.\\
$^{16}$Department of Astronomy, University of California, Berkeley, CA 94720, USA.\\
$^{17}$Department of Physics, Institute for Astronomy, ETH Z\"urich, CH-8093 Z\"urich, Switzerland.\\
$^{18}$Max-Planck-Institut f\"ur Astronomie, K\"onigstuhl 17, D-69117 Heidelberg, Germany.
}
\date{Received date: \today; accepted date: ***}

\begin{document}
\maketitle

\begin{abstract}
\CO~observations allow estimations of the gas content of molecular clouds, which trace the reservoir of cold gas fuelling star formation, as well as to determine extinction via \H~column density, \nh.
Here, we studied millimetric and optical properties at 26 supernovae (SNe) locations of different types in a sample of 23 nearby galaxies by combining molecular $^{12}$C$^{16}$O~({\it J} = 1 $\rightarrow$ 0)~resolved maps from the EDGE survey and optical Integral Field Spectroscopy from the CALIFA survey.
We found an even clearer separation between type II and type Ibc SNe in terms of molecular gas than what we found in the optical using H$\alpha$ emission as a proxy for current SF rate, which reinforces the fact that SNe Ibc are more associated with SF-environments. 
While \av~at SN locations is similar for SNe II and SNe Ibc, and higher compared to SNe Ia, \nh~is significantly higher for SNe Ibc than for SNe II and SNe Ia.
When compared to alternative extinction estimations directly from SN photometry and spectroscopy, we find that our SNe Ibc have also redder color excess but showed standard Na I D absorption pseudo-equivalent widths ($\sim$1 \AA).
In some cases we find no extinction when estimated from the environment, but high amounts of extinction when measured from SN observations, which suggests that circumstellar material or dust sublimation may be playing a role.
This work serves as a benchmark for future studies combining last generation millimeter and optical IFS instruments to reveal the local environmental properties of extragalactic SNe.
\end{abstract}


\section{Introduction}

It is now well established, both photometrically \citep{2012MNRAS.424.1372A,2012ApJ...759..107K} and spectroscopically \citep{2014A&A...572A..38G}, that stripped envelope supernovae (SESNe, types Ib, Ic and IIb) are more closely associated with star-forming regions than other core collapse supernovae (CCSNe) that retain most of  the content in their external layers before explosion (SNe II, historical types IIP, IIL, and IIn). 
Direct detection of SN progenitors in pre-explosion images is very rare and only a handful of objects have been reported, most of them being SN IIP progenitors \citep{2015PASA...32...16S}. However the study of SN environments have proved to be useful in putting constraints on SN progenitor scenarios.
While CCSNe are the final stage in the evolution of stars with zero-age main-sequence mass M$_{ZAMS}$ $>$ 8 \Msun~up to 25-30 \Msun, stars with lower masses ($<$ 8 \Msun) evolve to become white dwarfs of around 0.6 \Msun~on average \citep{2007MNRAS.375.1315K}, and only those in binary systems and that are able to accrete material from its companion reaching the right densities to explode as type Ia supernovae (SNe Ia).

Within the wide range of possible masses of CCSNe progenitors, if SESNe progenitors are single stars they are expected to come from those in the higher end of the mass range (a viable candidate being Wolf-Rayet stars; \citealt{2007ARA&A..45..177C}), and to be located in metal-richer environments responsible to get rid of their external layers by means of stellar metallicity driven winds \citep{2000ARA&A..38..613K,2005A&A...442..587V}\footnote{There is growing evidence that SESN progenitors are a mixture of single stars and binary systems. This picture stands in the case of single star progenitors, but also in the case of binaries since the mass transfer rate is also metallicity dependent (e.g. \citealt{2013ApJ...762...74B}).}. Assuming any initial mass function (IMF), just a few stars with higher masses are created in molecular clouds, and their evolution is much faster than their less massive counterparts.
This short lifetime causes the resulting SESNe to be found very close to the progenitor birth place.
The progenitors of the less massive types of CCSNe ($\sim$8 \Msun), SNe II, have lifetimes of around 40 Myr \citep{2000A&AS..141..371G}, and although they also explode close to where they were formed, the longer delay allows them to travel away from their birthplace.
Finally, the birthplace of SN Ia progenitor systems, whose lifetime ranges from a few hundreds of Myr to 10 Gyr \citep{2014ARA&A..52..107M}, have no direct connection with the location where the explosion occurred \citep{2014A&A...572A..38G,2016A&A...591A..48G}.

The relation between the SN type and the local star formation has been proved by analyzing the H$\alpha$ emission at the SN location (or at the nearest HII region), since it traces the content of atomic Hydrogen warm gas ionized by massive and young stars (mostly types O and B), and therefore it is a good proxy for on-going star formation ($\sim$10 Myr, \citealt{1998ARA&A..36..189K,2015A&A...584A..87C}). 
Besides the fact that CCSN progenitors are young stars and are found at locations with young stellar populations, it has also been found that SNe Ia explode at ubiquitous positions within their host galaxies given their longer lifetime. 
More accurate characterization of the parameters at SN positions can help us improve our knowledge about the mechanisms that produce different SN types, and it can also reduce the systematic uncertainties in the SN (both Ia and II) methods to estimate cosmological distances by introducing an environmental parameter in their brightness standardization \citep{2014A&A...568A..22B, 2015ApJ...815..121D}. 

Although the ratio of molecular (H$_2$) to atomic (H I) gas ratio decreases along the Hubble sequence from early- to late-type galaxies \citep{1989ApJ...347L..55Y,2014A&A...564A..66B}, cold gas (as opposed to neutral and ionized medium; H II) is still the dominant phase of the ISM in late-type galaxies \citep{1994ApJ...428..638S}.
Primordial gaseous material condensates into molecular clouds to then collapse into stars.
Star formation is fuelled by molecular gas, so it can be understood as a reservoir where new bursts of star formation may one day occur.  
The most abundant molecule in the universe is molecular Hydrogen, \H, but it is not directly observable in emission at the typical physical conditions present in the ISM of molecular clouds due to its symmetric structure which does not create a permanent electric dipole moment.  
Despite this situation, the second most abundant molecule in the cold ISM, carbon monoxide, is easily observed at the low temperatures of these clouds and has been proved to be a good proxy for \H~because \CO~and \H~are coextensive at solar metallicities \citep{1991ARA&A..29..581Y}. 
An explicit relation,  the \CO~to \H~conversion factor $X_{\CO}$, relates the measured \CO~intensity and the amount of \H~in a cloud \citep{1983QJRAS..24..267H,2013ARA&A..51..207B,2013tra..book.....W}.
Although $X_{\CO}$ might change with the different physical conditions characterizing the properties of the ISM, such as the metallicity and the CO luminosity,  
it can be constrained by studying its behavior in the Milky Way (MW) and other nearby galaxies.
Therefore, the amount of \CO~molecular gas can be used to estimate quantities related to the \H~gas, such as the molecular gas mass (M$_{H_2}$) and the \H~column density (\nh).

The atomic gas column density is in turn related to the optical extinction in the line-of-sight: usually denser \H~columns are found where larger amounts of dust are also present.
CO is therefore at the same time a proxy for the presence of \H, gas reservoir, and optical extinction. 
Measuring optical extinction and \H~column density at SN positions provides information about the insterstellar nature of the source producing the observed reddening, if it is present.
Photometric and spectroscopic methods have been developed to estimate extinction during few weeks after SN explosion, while the SN light is still visible and observable (e.g. \citealt{2010AJ....139..120F}). In some cases, a contribution may be caused by circumstellar material only visible at that time. This may be the case for SESNe, whose outer layers lost before explosion are located within a few pc of the progenitor star, although SNe IIn and some SNe Ia have been also associated with CSM interaction (e.g. \citealt{2007A&A...474..931P,2013ApJ...779...38P,2016MNRAS.458.2063K}).

Nowadays, no statistical studies of extragalactic SN locations have been performed in a different wavelength range that proves whether the same sequence found in the optical between different SN types and the strength of the star formation is also valid in millimiter wavelengths (but see \citealt{2013MNRAS.436.3464K} for UV/NIR).
This is one of the purposes of this paper.
Combining optical Integral Field Spectroscopy data from the Calar Alto Legacy Integral Field Area (CALIFA) survey and millimetric CO observations from the Extragalactic Database for Galaxy Evolution (EDGE) survey, we compare for the first time star formation and extinction proxies from the SN environment with those measured from SN photometry and spectroscopy. 

This paper is structured as follows:
In section \ref{sec:sam} we describe our sample and the two surveys, CALIFA and EDGE, where our data come from.
Then, in section \ref{sec:ana}, the parameters measured for the first time in this work from \CO~maps are presented.
In sections \ref{sec:mass}, \ref{sec:sfr}, and \ref{sec:ext} we present the local and global correlations of the optical and millimeter mass, star formation and extinction parameters, respectively.
In section \ref{sec:sn} we compiled photometric and spectroscopic data of a subsample of SNe, and compared our host galaxy extinction measurements to alternative methods from SN data itself.
Finally, in section \ref{sec:conc} we discuss our results and summarize our conclusions.

\begin{table*}\footnotesize
\caption{Properties of the 23 galaxies and their 26 SNe observed with both CALIFA and EDGE used in this work. The morphological galaxy type, redshift, and SN angular separation from the galaxy core are from the NASA/IPAC Extragalactic Database (NED),  SN type is obtained from the Asiago SN catalogue, and luminosity distance is determined assuming a flat $\Lambda$CDM cosmological model with H$_0$=70 and $\Omega_M$=0.27.}

\label{tab:sam}
\begin{center}
\begin{tabular}{llcclcc}
\hline\hline
Galaxy&Morphology&\multicolumn{1}{c}{z}& D$_L$ &SN name&Type&\multicolumn{1}{c}{Separation}\\
      &          &                     & [Mpc]&    & &\multicolumn{1}{c}{[arcsec]}  \\ 
\hline
NGC 0523 &pec           &0.015871&  68.0 & 2001en     &Ia    &26.6\\
NGC 2347 &(R')SA(r)b?   &0.014747&  63.2 & 2001ee     &II    &12.7\\
UGC 04132&Sbc           &0.017409&  74.7 & 2005en     &II    & 9.5\\  
         &              &        &       & 2005eo     &Ic    &29.8\\
         &              &        &       & 2014ee     &IIn   &24.9\\
NGC 2623 &pec           &0.018509&  79.5 & 1999gd     &Ia    &17.9\\
NGC 2906 &Scd?          &0.007138&  30.4 & 2005ip     &IIn   &14.3\\
NGC 2916 &SA(rs)b?      &0.012442&  53.2 & 1998ar     &II    &42.3\\
NGC 3687 &(R')SAB(r)bc? &0.008362&  35.6 & 1989A      &Ia    &33.6\\
NGC 3811 &SB(r)cd?      &0.010357&  44.2 & 1969C      &Ia    &33.6\\                
         &              &        &       & 1971K      &IIP   &35.9\\                     
NGC 4210 &SB(r)b        &0.009113&  38.9 & 2002ho     &Ic    &17.8\\
NGC 4644 &SBb?          &0.016501&  70.8 & 2007cm     &IIn   &25.4\\  
NGC 4961 &SB(s)cd       &0.008456&  36.0 & 2005az     &Ic    & 9.7\\  
NGC 5000 &SB(rs)bc      &0.018706&  80.4 & 2003el     &Ic    &17.2\\  
UGC 08250&Scd?          &0.017646&  75.8 & 2013T      &Ia    &34.5\\
NGC 5056 &Scd?          &0.018653&  80.2 & 2005au     &II    &20.7\\
NGC 5480 &SA(s)c?       &0.006191&  26.3 & 1988L      &Ib    &10.7\\
NGC 5682 &SB(s)b        &0.007581&  32.3 & 2005ci     &II    & 7.0\\
NGC 5732 &Sbc           &0.012502&  53.5 & ASASSN-14jf&II    &20.0\\
NGC 5980 &S             &0.013649&  58.4 & 2004ci     &II    & 8.6\\  
UGC 10331&S pec         &0.014914&  63.9 & 2011jg     &IIb   &24.5\\  
NGC 6063 &Scd?          &0.009500&  40.5 & 1999ac     &Ia-pec&38.3\\
NGC 6146 &E?            &0.029420& 127.5 & 2009fl     &Ia    &13.8\\
NGC 6186 &(R')SB(s)a    &0.009797&  41.8 & 2011gd     &Ib    & 2.9\\  
UGC 10123&Sab           &0.012575&  53.8 & 2014cv     &IIP   & 6.0\\
\hline
\end{tabular}
\end{center}
\end{table*}



\section{Galaxy sample} \label{sec:sam}

The 23 galaxies used in this work were selected from the subsample of 115 nearby galaxies observed with optical Integral Field Spectroscopy (IFS) by the CALIFA survey \citep{2012A&A...538A...8S,2016A&A...591A..48G}, that hosted a discovered SN of any type, and that have also millimeter spatially resolved data obtained by the EDGE survey available.
In Table \ref{tab:sam} we list the main properties of the 23 galaxies and of the 26 SNe these galaxies hosted. 
Below, we briefly describe both surveys.

\subsection{CALIFA survey} 

The Calar Alto Legacy Integral Field Area (CALIFA) Survey was performed with the purpose of obtaining spatially resolved spectroscopic information of $\sim$600 galaxies in the near universe ($0.005 <z < 0.03$), using the Potsdam Multi Aperture Spectrograph \citep[PMAS][]{2005PASP..117..620R} Integral Field Unit (IFU) in the PPAk mode \citep{2004AN....325..151V, 2006PASP..118..129K}, mounted to the 3.5m telescope of the Centro Astron\'omico Hispano-Alem\'an (CAHA) at the Calar Alto Observatory, Almer\'ia.
These $\sim$600 galaxies were selected from the {\it CALIFA Mother Sample} that was constructed from galaxies in the Seventh Data Release of the Sloan Digital Sky Survey \citep[SDSS DR7, ][]{2009ApJS..182..543A} following these criteria:
(i) galaxies must show recession velocities within the redshift range $0.005 < z < 0.03$;
(ii) they must have an angular isophotal diameter in the range $45< D_{25}< 80$ arcsec, to maximize the use of the large FoV ($74\times65$ arcsec) of the instrument; and
(iii) they must be located at $\delta>$-7 deg for galaxies in the North Galactic hemisphere, to ensure good visibility from the observatory.

A detailed description of the survey is presented in \cite{2012A&A...538A...8S}, and the galaxy sample characterization in \cite{2014A&A...569A...1W}. The main data products of CALIFA survey are various properties and kinematic maps of the stellar populations and ionized gas, which can be found in its Third Data Release (DR3, \citealt{2016arXiv160402289S}).

\begin{table*}\footnotesize
\caption{Values of the parameters estimated in the total integrated signal from EDGE CO 0th-moment maps and CALIFA optical spectra.}
\label{tab:tot}
\begin{center}
  \begin{tabular}{lcccccccc}
    \hline
    \hline
Galaxy name  &\multicolumn{1}{c}{ Beam area} &\multicolumn{1}{c}{$S_{CO}\Delta v$}&\multicolumn{1}{c}{M$_{\rm mol}$}&\multicolumn{1}{c}{N(H$_2$)}            &\multicolumn{1}{c}{$F_{H\alpha}$}          &\multicolumn{1}{c}{M$_*$}      &\multicolumn{1}{c}{A$_V$}\\
             &\multicolumn{1}{c}{ arcsec$^2$}&\multicolumn{1}{c}{Jy~km~s$^{-1}$}  &\multicolumn{1}{c}{log (\Msun)}  &\multicolumn{1}{c}{$10^{21} ~$cm$^{-2}$}& 10$^{-13}$~erg~s$^{-1}$~cm$^2$~\AA$^{-1}$ &\multicolumn{1}{c}{log (\Msun)}&\multicolumn{1}{c}{mag}  \\
             & (a)                           & (b)                                & (c)                             & (d)                                    & (e)                                       & (f)                           & (g)                     \\
\hline
NGC 0523     &   19.8  &  94.35$\pm$5.51 &  9.66 & 2.10$\pm$0.07 & 3.49 $\pm$ 0.30 & 10.72 & 1.19 $\pm$ 0.18 \\ 
NGC 2347     &   26.3  &  86.33$\pm$4.65 &  9.55 & 1.17$\pm$0.04 & 6.18 $\pm$ 0.25 & 10.79 & 0.50 $\pm$ 0.07 \\ 
UGC 04132    &   24.6  & 179.25$\pm$6.02 & 10.02 & 2.76$\pm$0.05 & 5.41 $\pm$ 0.25 & 10.93 & 0.95 $\pm$ 0.14 \\ 
NGC 2623     &   22.5  & 125.81$\pm$3.06 &  9.92 & 4.53$\pm$0.06 & 0.70 $\pm$ 0.06 & 10.56 & 1.38 $\pm$ 0.21 \\ 
NGC 2906     &   29.8  &  87.66$\pm$6.32 &  8.93 & 1.10$\pm$0.04 & 5.23 $\pm$ 0.22 & 10.36 & 0.81 $\pm$ 0.12 \\ 
NGC 2916     &   19.9  &  38.44$\pm$5.34 &  9.05 & 0.77$\pm$0.07 & 5.58 $\pm$ 0.24 & 10.62 & 0.53 $\pm$ 0.08 \\ 
NGC 3687     &   23.9  &  $<$27.86       &$<$8.41&$<$0.13        & 3.36 $\pm$ 0.11 & 10.23 & 0.44 $\pm$ 0.07 \\ 
NGC 3811     &   22.8  &  93.71$\pm$6.47 &  9.28 & 1.31$\pm$0.05 & 6.62 $\pm$ 0.35 & 10.44 & 0.81 $\pm$ 0.12 \\ 
NGC 4210     &   11.1  &  46.46$\pm$5.59 &  8.87 & 0.82$\pm$0.07 & 3.74 $\pm$ 0.12 & 10.25 & 0.47 $\pm$ 0.07 \\ 
NGC 4644     &   23.9  &  29.71$\pm$3.29 &  9.19 & 0.77$\pm$0.05 & 1.15 $\pm$ 0.06 & 10.41 & 0.63 $\pm$ 0.09 \\ 
NGC 4961     &   21.9  &  18.47$\pm$3.55 &  8.40 & 0.85$\pm$0.09 & 4.69 $\pm$ 0.20 & 9.79  & 0.25 $\pm$ 0.04 \\ 
NGC 5000     &   19.4  &  41.77$\pm$3.89 &  9.45 & 1.29$\pm$0.06 & 1.71 $\pm$ 0.11 & 10.70 & 1.14 $\pm$ 0.17 \\ 
UGC 08250    &   21.6  &  $<$19.76       &$<$8.91&$<$0.05        & 0.82 $\pm$ 0.06 & 10.03 & 0.64 $\pm$ 0.10 \\ 
NGC 5056     &   28.1  &  41.36$\pm$3.82 &  9.44 & 0.82$\pm$0.04 & 4.72 $\pm$ 0.19 & 11.03 & 0.41 $\pm$ 0.06 \\ 
NGC 5480     &   18.1  & 109.91$\pm$7.22 &  8.90 & 1.39$\pm$0.05 &10.60 $\pm$ 0.48 & 9.99  & 0.88 $\pm$ 0.13 \\ 
NGC 5682     &   19.9  &  $<$25.39       &$<$8.28&$<$0.05        & 2.63 $\pm$ 0.16 & 9.40  & 0.30 $\pm$ 0.05 \\ 
NGC 5732     &   25.6  &  22.10$\pm$3.70 &  8.82 & 0.68$\pm$0.10 & 2.42 $\pm$ 0.12 & 9.89  & 0.31 $\pm$ 0.05 \\ 
NGC 5980     &   21.9  & 136.56$\pm$5.20 &  9.69 & 1.90$\pm$0.05 & 6.45 $\pm$ 0.25 & 10.73 & 0.78 $\pm$ 0.12 \\ 
UGC 10331    &   30.0  &  $<$29.64       &$<$8.94&$<$0.24        & 4.43 $\pm$ 0.35 & 10.01 & 0.78 $\pm$ 0.12 \\ 
NGC 6063     &   26.0  &  $<$28.53       &$<$8.53&$<$0.11        & 2.47 $\pm$ 0.08 & 10.13 & 0.46 $\pm$ 0.07 \\ 
NGC 6146     &   23.4  &  $<$19.49       &$<$9.35&$<$0.05        & 0.25 $\pm$ 0.01 & 11.27 & 0.01 $\pm$ 0.10 \\ 
NGC 6186     &   22.9  & 154.77$\pm$6.78 &  9.45 & 3.09$\pm$0.07 & 4.45 $\pm$ 0.39 & 10.49 & 1.11 $\pm$ 0.17 \\ 
UGC 10123    &   19.2  & 101.43$\pm$5.29 &  9.49 & 2.16$\pm$0.07 & 1.88 $\pm$ 0.12 & 10.51 & 1.40 $\pm$ 0.21 \\ 
\hline
\end{tabular}
\end{center}
(a) EDGE beam area. 
(b) CO integrated line flux. 
(c) Molecular mass.
(d) H$\alpha$ flux.
(e) H$_2$ column density.
(f) Optical stellar mass.
(g) Optical visual extinction.
\end{table*}

\subsection{EDGE survey}

The Extragalactic Database for Galaxy Evolution (EDGE) survey was designed from the outset to complement a subsample of galaxies with CALIFA IFS data in the optical with millimeter interferometric observations using the Combined Array for Millimeterwave Astronomy (CARMA), an array of 23 Radio telescopes that was located at the Inyo mountains, California\footnote{CARMA ceased operations and was decomissioned in April 2015.}, but with an additional bias toward infrared (IR) bright galaxies in light of the well-known correlation between IR and CO luminosity.
EDGE survey is the first major, resolved CO survey matched to an IFU survey. The full survey description will be presented elsewhere (Bolatto et al. in prep.), but we provide here a brief summary. It is anticipated that the EDGE data products will be publicly available after the survey description is finalized.

The observations were carried out between November 2014 and April 2015, using half-beam spaced 7-point hexagonal mosaics to optimize extended flux recovery and uniformity of sensitivity over the central 1\arcmin\ diameter region. A first selection of 177 galaxies from the CALIFA observed galaxies were taken in the {\it J} = 1 $\rightarrow$ 0 transition of $^{12}$C$^{16}$O at $\sim$2.6 mm (115.27 GHz) using the E-array configuration (40 min of integration, $8''$ resolution, and baseline of 8.5-66 meters).
A subset of these galaxies (126) that were judged promising (or that otherwise fit the available observing schedule) were followed up in the D-array configuration to achieve higher angular resolution (typically 3.5 hours of integration, $4.5''$ resolution equivalent to $\sim 1-2$ kpc, and baseline of 11-148 meters).

For reasons of observing efficiency galaxies were divided in three groups using their optical redshift, and each group was observed with a fixed tunning and correlator setup. The correlator had five 250 MHz windows covering the region were the CO {\it J} = 1 $\rightarrow$ 0 transition was expected, producing data at 3.4 km\,s$^{-1}$ resolution. The final maps combined the E and D array observations and were produced with 20 km\,s$^{-1}$ velocity resolution using a standard pipeline written in Miriad. The typical angular resolution attained is 4.5\arcsec, and the typical sensitivity is 30 mK in 20 km\,s$^{-1}$. Because of changing integration times and observing conditions there is a distribution of sensitivities, but 95\% of the galaxies fall in the 20-39 mK interval.
The survey is thus sensitive to an H$_2$ surface density of $\sim$10 \Msun~pc$^{-2}$ (averaged over a $\sim$1.5 kpc scale), assuming a standard CO-to-H$_2$ conversion factor. Of the 126 galaxies observed in both D and E arrays, 105 display a CO\footnote{Hereafter, we will refer to the ground rotational transition {\it J} = 1 $\rightarrow$ 0 of $^{12}$C$^{16}$O as simply CO.} peak temperature at the 5$\sigma$ or higher level and thus are considered secure detections. 
Twenty-three of these 126 galaxies were also in the CALIFA SN host galaxy sample, and therefore selected for this study.

\section{Analysis}\label{sec:ana}

Here we describe how we measured the required parameters for the analysis.
We employed our Python codes to work with the information contained in both datasets and estimated the parameters detailed below.
Both total and local measurements of the parameters described in this section are reported in Tables \ref{tab:tot} and \ref{tab:loc}, respectively.
For EDGE data where no measurement was possible we report upper limits for non-detections, which were estimated as 3 times the error at the position and give an idea of how bright a line of $\lesssim$200 km s$^{-1}$ width could be buried in the noise.

\subsection{Millimeter parameters from EDGE} 

Reduction and analysis of CARMA data was performed with the EDGE pipeline (Wong et al. in prep.).
The main data products are spatially resolved maps of the first three moments (0th, 1st, and 2nd), which correspond to integrated line intensity, mean velocity, and velocity dispersion of the CO molecular line, respectively. The 0th moment maps have K km s$^{-1}$ units, corresponding to integrating the brightness temperature along the spectral axis in km s$^{-1}$.

\subsubsection{CO integrated line flux} 

\begin{table*}\fontsize{7.2}{8.64}\selectfont
\caption{Values of the parameters estimated at the SN location from EDGE CO 0th-moment maps, CALIFA optical spectra, and SN photometry and spectroscopy.}
\label{tab:loc}
\begin{center}
  \begin{tabular}{lccccccccc}
    \hline
    \hline

SN name& Type   & $S_{CO}\Delta v$  & N(H$_2$)          &\multicolumn{1}{c}{$F_{H\alpha}$}          & A$_V$           & A$_V^*$ &  E(B-V)$_{Na D}$  &  E(B-V)$_c$       & E(B-V)$_{\rm color~curve}$\\
       &        & Jy~km~s$^{-1}$    &$10^{21}~$cm$^{-2}$& 10$^{-16}$~erg~s$^{-1}$~cm$^2$~\AA$^{-1}$ &mag              & mag     & mag               &mag                &mag                        \\
       &        & (a)               & (b)               & (c)                                       & (d)             & (e)     & (f)               & (g)               & (h)                       \\
 \hline                                                                                                                                                                                             
2001en & Ia     & 0.026 $\pm$ 0.011 & 0.54 $\pm$ 0.24   & 0.59 $\pm$ 0.08                           & $<$0.01         & 0.22    & 0.170 $\pm$ 0.013 & 0.010 $\pm$ 0.013 & 0.079 $\pm$ 0.016         \\ 
1999gd & Ia     & $<$0.016          & $<$0.33           & 0.30 $\pm$ 0.05                           & $<$0.01         & 0.15    &      $>$2.000     & 0.305 $\pm$ 0.033 & 0.417 $\pm$ 0.083         \\ 
1989A  & Ia     & $<$0.038          & $<$0.79           & 0.48 $\pm$ 0.07                           & 0.27 $\pm$ 0.04 & 0.06    &        ---        & 0.054 $\pm$ 0.033 & 0.003 $\pm$ 0.001         \\ 
1969C  & Ia     & $<$0.009          & $<$0.19           & 0.41 $\pm$ 0.06                           & 0.06 $\pm$ 0.01 & $<$0.01 &        ---        &        ---        & 0.251 $\pm$ 0.050         \\ 
2013T  & Ia     & $<$0.032          & $<$0.67           & 0.32 $\pm$ 0.05                           & $<$0.01         & $<$0.01 &        ---        &        ---        &        ---                \\ 
1999ac & Ia-pec & $<$0.047          & $<$0.98           & 0.40 $\pm$ 0.06                           & $<$0.01         & $<$0.01 & 0.020 $\pm$ 0.005 & 0.061 $\pm$ 0.013 & 0.088 $\pm$ 0.018         \\ 
2009fl & Ia     & $<$0.023          & $<$0.48           & 0.01 $\pm$ 0.00                           & $<$0.01         & $<$0.01 &        ---        &        ---        &        ---                \\ 
 \hline                                                                                                                                                                                             
2001ee & II     & 0.049 $\pm$ 0.021 & 1.02 $\pm$ 0.44   & 2.11 $\pm$ 0.31                           & 0.80 $\pm$ 0.12 & 0.06    &        ---        &       ---         &        ---                \\ 
2005en & II     & 0.024 $\pm$ 0.013 & 0.50 $\pm$ 0.28   & 7.08 $\pm$ 1.02                           & 1.10 $\pm$ 0.16 & 0.30    & 0.225 $\pm$ 0.050 &       ---         &        ---                \\ 
2014ee & IIn    & $<$0.026          & $<$0.54           & 3.45 $\pm$ 0.50                           & 1.13 $\pm$ 0.17 & 0.08    & 0.160 $\pm$ 0.026 &       ---         &        ---                \\ 
2005ip & IIn    & 0.009 $\pm$ 0.007 & 0.19 $\pm$ 0.12   & 6.31 $\pm$ 0.83                           & 2.27 $\pm$ 0.34 & 0.09    & 0.251 $\pm$ 0.067 &       ---         & 0.431 $\pm$ 0.086         \\ 
1998ar & II     & $<$0.045          & $<$0.94           & 0.36 $\pm$ 0.04                           & $<$0.01         & $<$0.01 &        ---        &       ---         &        ---                \\ 
1971K  & IIP    & $<$0.035          & $<$0.73           & 0.24 $\pm$ 0.03                           & 0.60 $\pm$ 0.09 & $<$0.01 &        ---        &       ---         &        ---                \\ 
2007cm & IIn    & $<$0.022          & $<$0.46           & 0.41 $\pm$ 0.06                           & 0.08 $\pm$ 0.01 & 0.10    &        ---        &       ---         &        ---                \\ 
2005au & II     & 0.021 $\pm$ 0.010 & 0.44 $\pm$ 0.22   & 7.84 $\pm$ 1.20                           & 0.46 $\pm$ 0.07 & 0.20    & 0.220 $\pm$ 0.053 &       ---         & 0.421 $\pm$ 0.084         \\ 
2005ci & II     & $<$0.032          & $<$0.67           & 3.05 $\pm$ 0.45                           & 0.49 $\pm$ 0.07 & 0.21    & 0.383 $\pm$ 0.075 &       ---         & 0.087 $\pm$ 0.017         \\ 
ASASSN-14jf& II & $<$0.029          & $<$0.60           & 0.99 $\pm$ 0.14                           & 0.28 $\pm$ 0.04 & 0.73    &        ---        &       ---         &        ---                \\ 
2004ci & II     & 0.010 $\pm$ 0.008 & 0.21 $\pm$ 0.14   & 3.74 $\pm$ 0.54                           & 0.62 $\pm$ 0.09 & 0.12    & 0.164 $\pm$ 0.018 &       ---         &        ---                \\ 
2014cv & IIP    & 0.409 $\pm$ 0.027 & 8.50 $\pm$ 0.56   & 2.46 $\pm$ 0.34                           & 1.98 $\pm$ 0.30 & 0.30    &        ---        &       ---         &        ---                \\ 
 \hline                                                                                                                                                                                             
2005eo & Ic     & 0.101 $\pm$ 0.028 & 2.10 $\pm$ 0.58   & 4.25 $\pm$ 0.64                           & 0.89 $\pm$ 0.13 & 0.29    & 0.375 $\pm$ 0.034 &       ---         & 0.754 $\pm$ 0.151         \\ 
2002ho & Ic     & 0.045 $\pm$ 0.025 & 0.94 $\pm$ 0.53   & 1.38 $\pm$ 0.21                           & 0.25 $\pm$ 0.04 & 0.09    &        ---        &       ---         &        ---                \\ 
2005az & Ic     & 0.023 $\pm$ 0.010 & 0.48 $\pm$ 0.22   & 8.51 $\pm$ 1.27                           & 0.56 $\pm$ 0.08 & 0.14    & 0.151 $\pm$ 0.014 &       ---         & 0.694 $\pm$ 0.139         \\ 
2003el & Ic     & 0.012 $\pm$ 0.007 & 0.25 $\pm$ 0.12   & 1.15 $\pm$ 0.18                           & 2.32 $\pm$ 0.35 & 0.09    & 0.244 $\pm$ 0.048 &       ---         &        ---                \\ 
1988L  & Ib     & 0.183 $\pm$ 0.034 & 3.80 $\pm$ 0.71   & 6.33 $\pm$ 0.93                           & 1.12 $\pm$ 0.17 & 0.04    & 1.188 $\pm$ 0.203 &       ---         &        ---                \\ 
2011jg & IIb    & $<$0.034          & $<$0.71           & 1.56 $\pm$ 0.23                           & 0.89 $\pm$ 0.13 & 0.23    &        ---        &       ---         &        ---                \\ 
2011gd & Ib     & 0.796 $\pm$ 0.030 &16.55 $\pm$ 0.62   &26.34 $\pm$ 3.64                           & 1.42 $\pm$ 0.21 & 0.19    &        ---        &       ---         &        ---                \\ 
\hline
\end{tabular}
\end{center}
(a) CO flux density.
(b) H$_2$ column density.
(c) H$\alpha$ flux.
(d) Optical visual gas extinction.
(e) Optical visual stellar extinction.
(f) B-V color excess measured from Na D pseudoequivalent width.
(g) B-V color excess measured from SN Ia color parameter.
(h) B-V color excess measured from SN color curves.
\end{table*}

The 0th-moment 2D maps contain the CO integrated line intensity measured along the FoV of the galaxies in units of Kelvin~km~s$^{-1}$, integrated over the beam size. 
Following \cite{2013ARA&A..51..207B}, the CO integrated line flux (in Jy km s$^{-1}$) can be measured from the EDGE maps from
\begin{equation}
S_{CO} \approx 10.9 \times 10^{-3}~J^2~\theta^2~T_J,
\end{equation}
where $T_J$ is the integrated brightness temperature, J=1 for \CO~(1$\rightarrow$ 0) transition, and $\theta$ is the beam solid angle listed in Table 2.

We estimated the total CO integrated line flux by summing up the signal over each map.
The CO emission is formally detected at some position within the observed galaxy extension in 17 of the 23 galaxies. On six cases (NGC 3687, UGC 08250, NGC 5682, UGC10331, NGC 6063, and NGC 6146) the line is not detected and only upper limits on the presence of cold gas are reported.

In order to obtain the CO integrated line flux at the SN position, we determined their coordinates in the 0th-moment maps by using the astrometry reported in the data products.
The intensity in the maps at that position resulted from averaging over a Gaussian elliptical beam of a few arcseconds, with variable size depending on the galaxy, with semi-major axes from 3.33" for NGC 4210 to 6.5" for UGC10331.
At half of the SN positions (13 over 26) the CO emission was also below the detection limit of the instrument, and only upper limits are reported.

\subsubsection{Molecular mass}

The total molecular mass in a galaxy can be estimated by calculating first the CO luminosity, L$_{CO}$, which relates to the observed integrated line flux in galaxies via
\begin{equation} 
L_{CO} = 2453 S_{CO}\Delta v D_L^2 / (1 + z),
\end{equation}
where $S_{CO}\Delta v$ is the integrated line flux measured above, in Jy km s$^{-1}$, $D_L$ is the luminosity distance to the galaxy in Mpc, $z$ is the redshift, and L$_{CO}$ has units of K km s$^{-1}$ pc$^2$.
The total molecular mass, in units of M$_{\odot}$, is simply the CO luminosity times $\alpha_{CO}$, the mass-to-light ratio
\begin{equation}
M_{mol} = \alpha_{CO}~L_{CO}.
\end{equation}
For $\alpha_{CO}$ we use a value of 4.3 M$_{\odot}$ (K km s$^{-1}$ pc$^{-2})^{-1}$ which corresponds to a CO-to-H$_2$ conversion factor X$_{CO}$ of $2 \times 10^{20}$ cm$^{-2}$ (K km s$^{-1})^{-1}$, and is generally assumed to be representative of galaxy disk environments similar to those encountered in the solar neighborhood and most ``normal'' galaxy disks in terms of metallicity and local radiation field \citep[see the extensive discussion in][]{2013ARA&A..51..207B}.
Given that our galaxies have integrated metallicities close to the solar values (8.3 to 8.6 dex, \citealt{2016A&A...591A..48G}) our selection of $X_{CO}$ seems justified.
Note that $M_{{\rm mol}}$ includes a 36\% correction for the contribution of heavy elements by mass, basically Helium. 

\begin{figure*}
\centering
\includegraphics[trim=0.2cm 0.4cm 0.52cm 0cm, clip=true,width=0.33\textwidth]{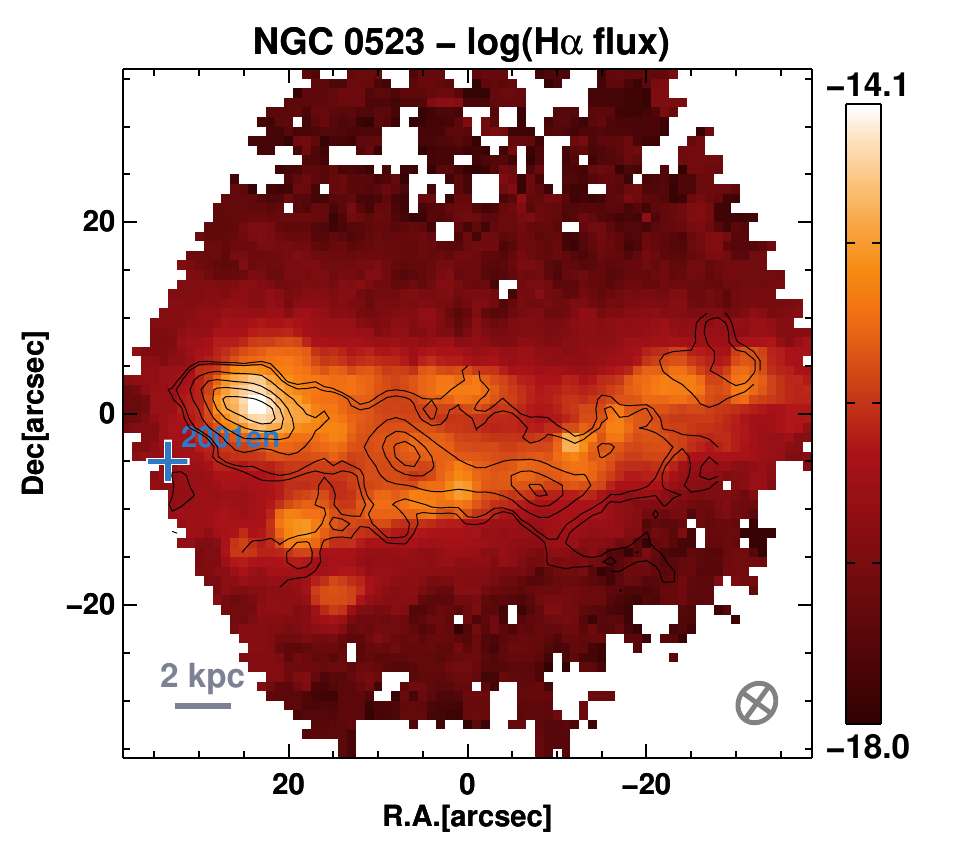}
\includegraphics[trim=0.52cm 0.4cm 0.2cm 0cm, clip=true,width=0.33\textwidth]{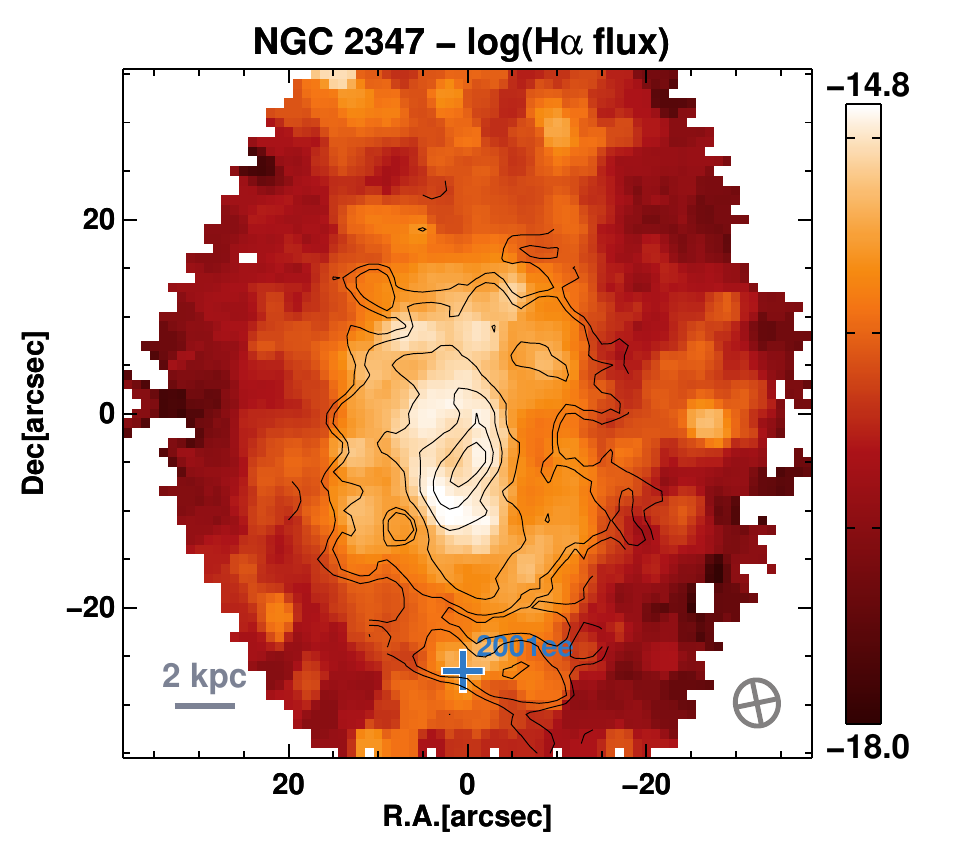}
\includegraphics[trim=0.52cm 0.4cm 0.2cm 0cm, clip=true,width=0.33\textwidth]{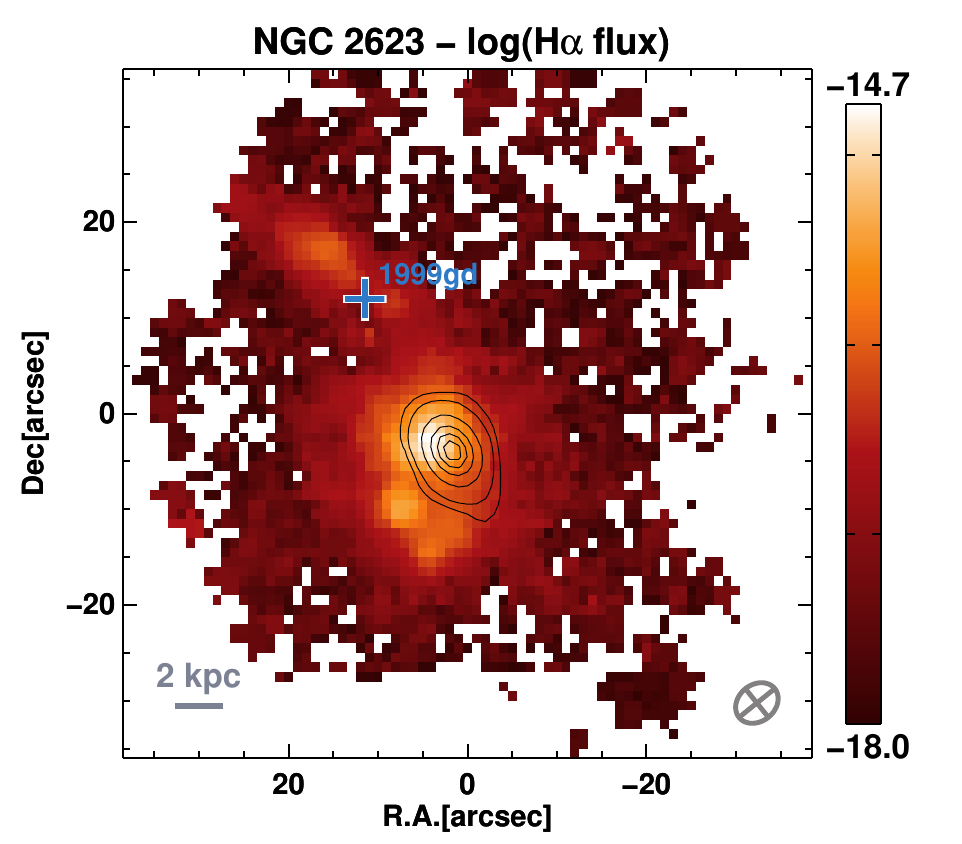}
\includegraphics[trim=0.2cm 0.4cm 0.52cm 0cm, clip=true,width=0.33\textwidth]{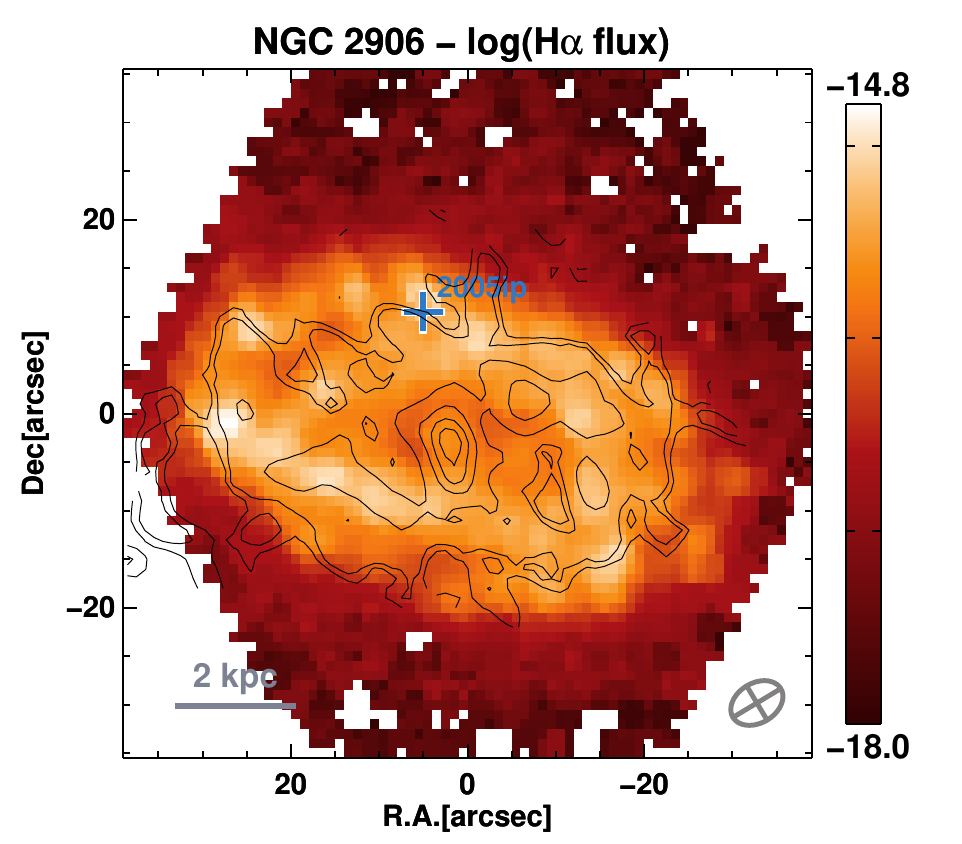}
\includegraphics[trim=0.52cm 0.4cm 0.2cm 0cm, clip=true,width=0.33\textwidth]{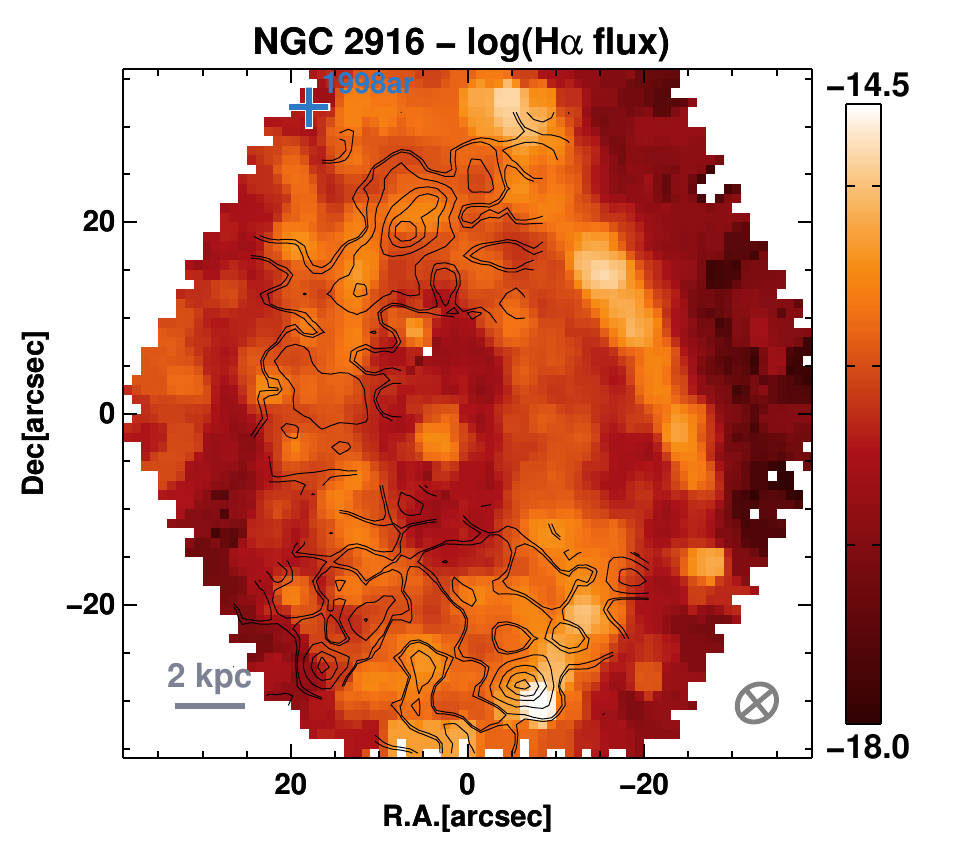}
\includegraphics[trim=0.52cm 0.4cm 0.2cm 0cm, clip=true,width=0.33\textwidth]{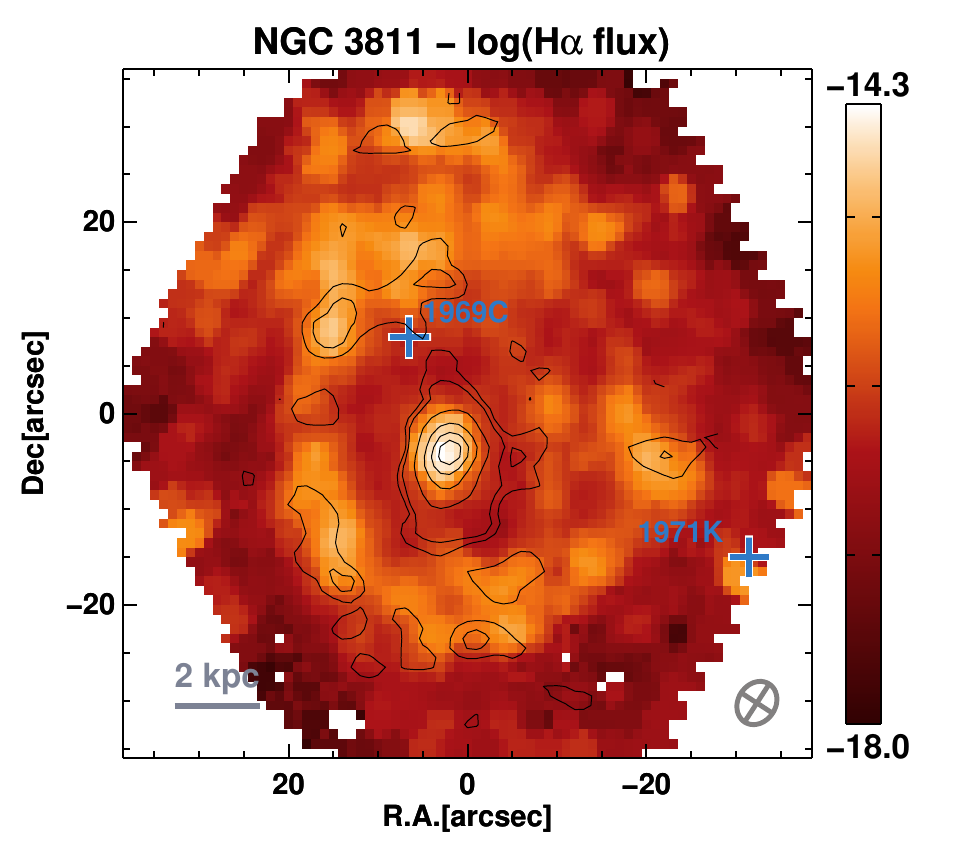}
\includegraphics[trim=0.2cm 0.4cm 0.52cm 0cm, clip=true,width=0.33\textwidth]{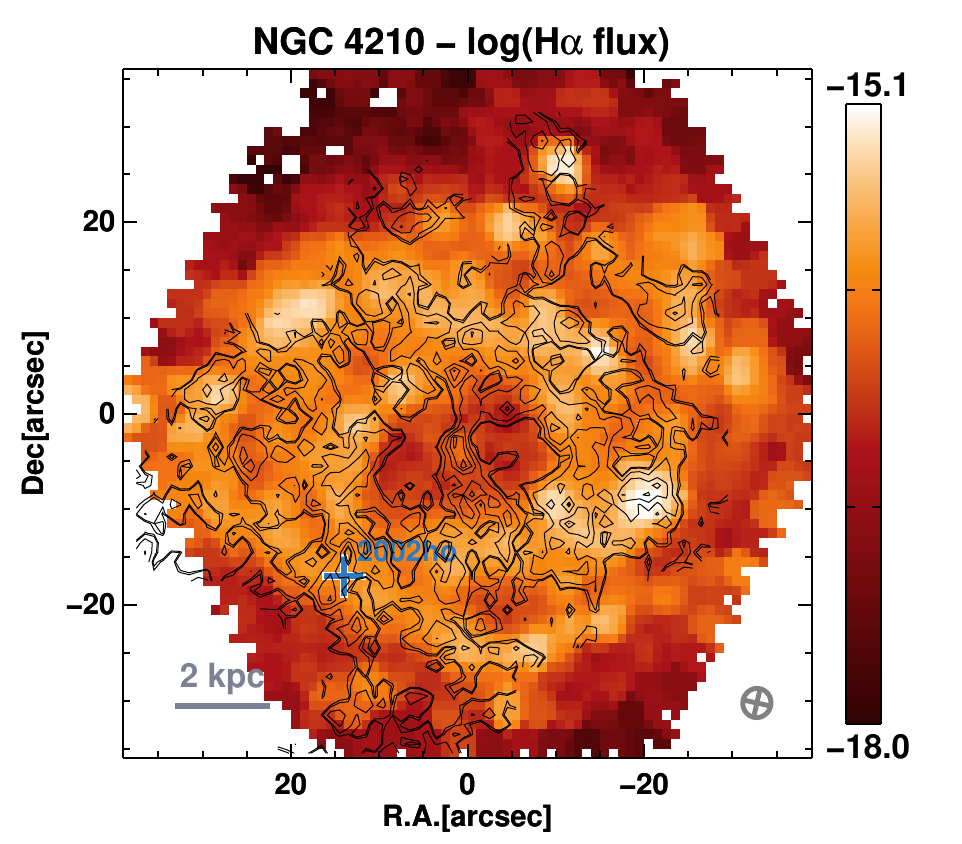}
\includegraphics[trim=0.52cm 0.4cm 0.2cm 0cm, clip=true,width=0.33\textwidth]{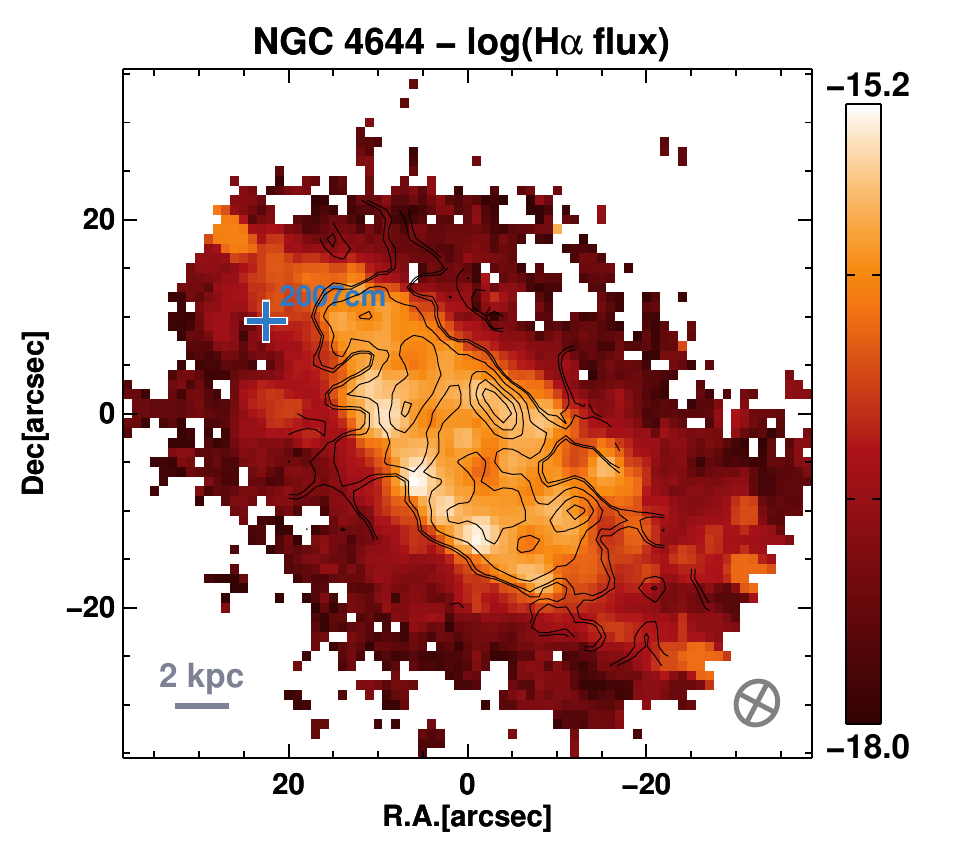}
\includegraphics[trim=0.52cm 0.4cm 0.2cm 0cm, clip=true,width=0.33\textwidth]{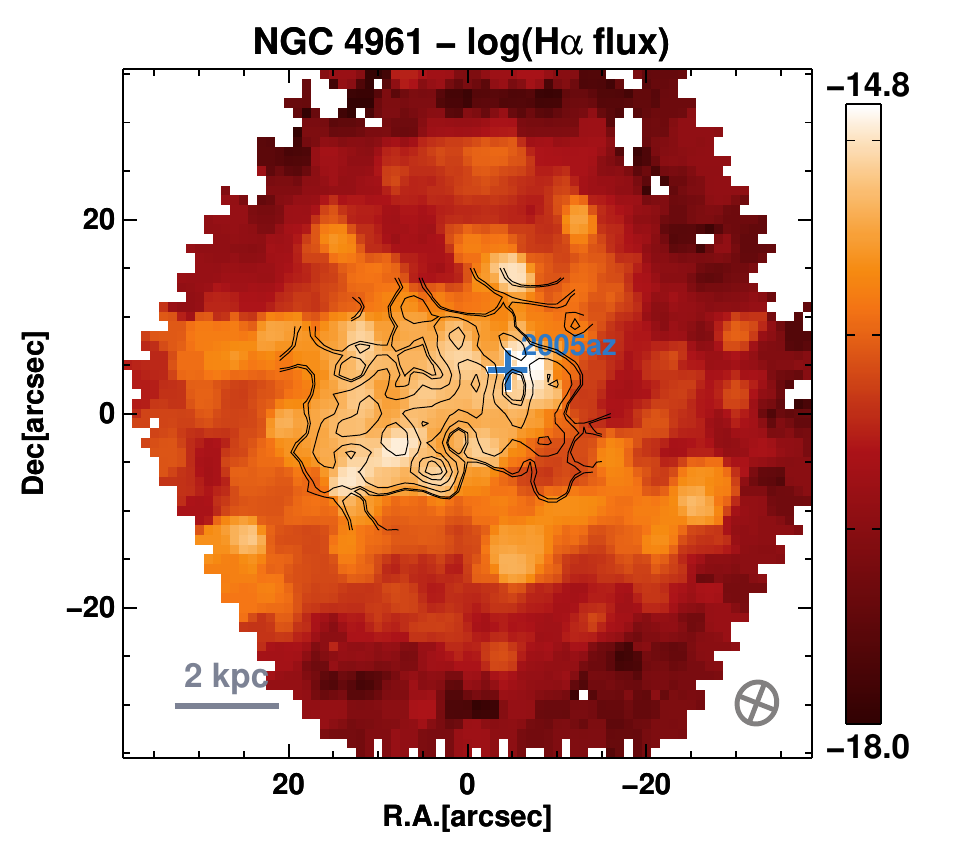}
\includegraphics[trim=0.2cm 0.1cm 0.52cm 0cm, clip=true,width=0.33\textwidth]{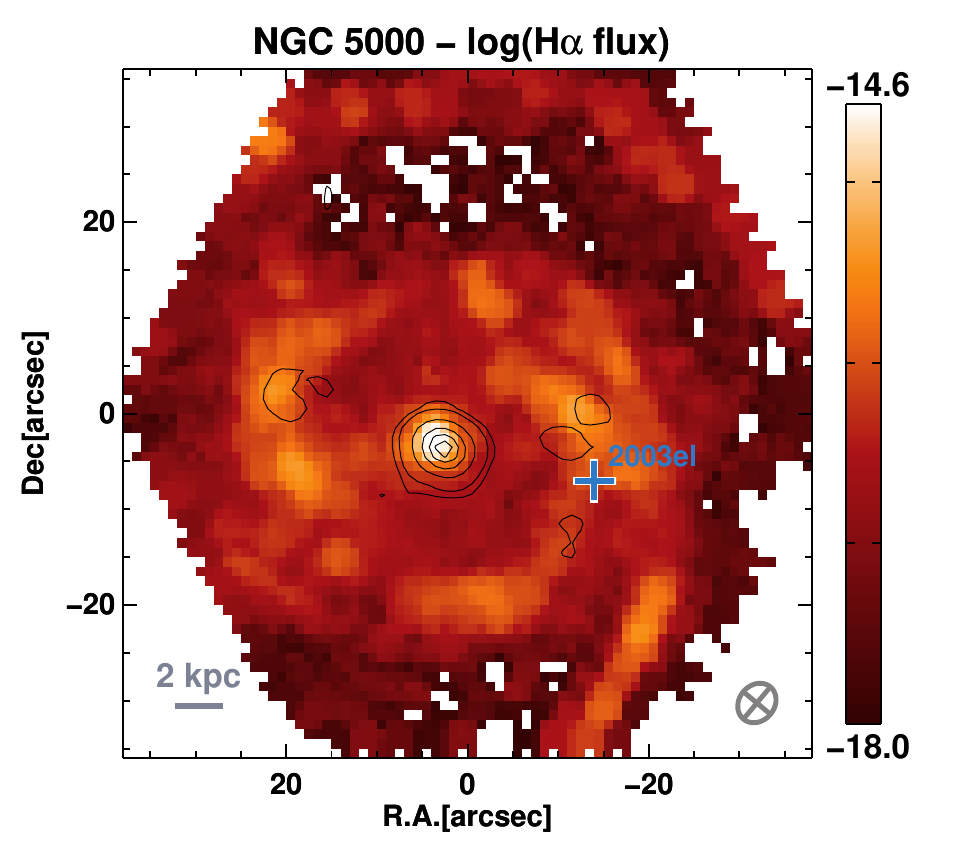}
\includegraphics[trim=0.52cm 0.1cm 0.2cm 0cm, clip=true,width=0.33\textwidth]{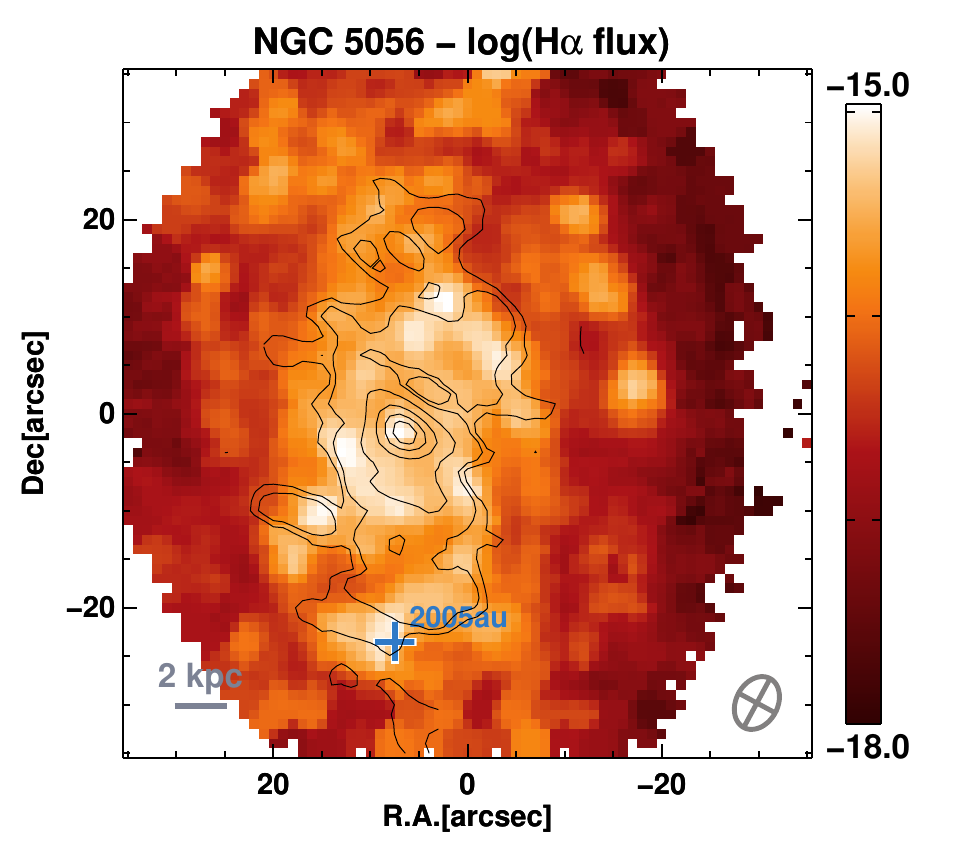}
\includegraphics[trim=0.52cm 0.1cm 0.2cm 0cm, clip=true,width=0.33\textwidth]{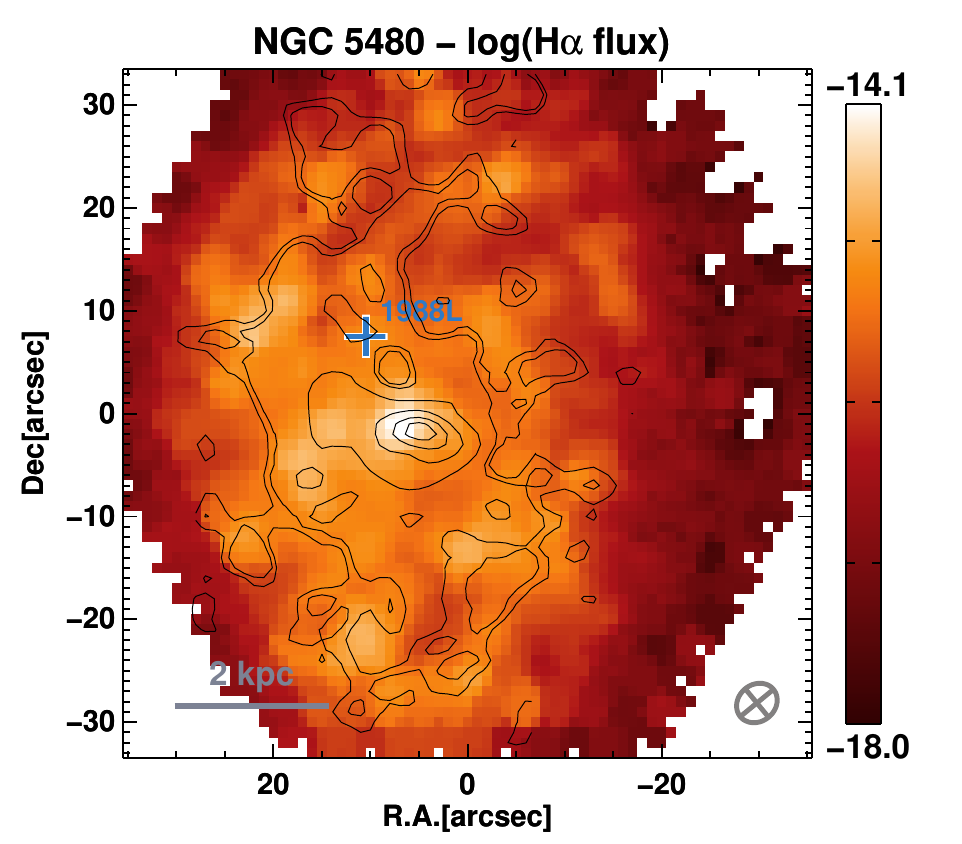}
\caption{$H\alpha$ emission maps for the 12 galaxies studied in this work with $CO$ intensities superposed in contour levels. Each contour represents the 5, 10, 30, 55, 75, and 90\% level, respectively, with respect the maximum emission. The CARMA beam size for each map is shown in the lower right corner of each box. SN positions are marked with green crosses.}
\label{fig:maps}
\end{figure*}
\renewcommand{\thefigure}{\arabic{figure} (Cont.)}
\addtocounter{figure}{-1}
\begin{figure*}
\centering
\includegraphics[trim=0.2cm 0.1cm 0.52cm 0cm, clip=true,width=0.33\textwidth]{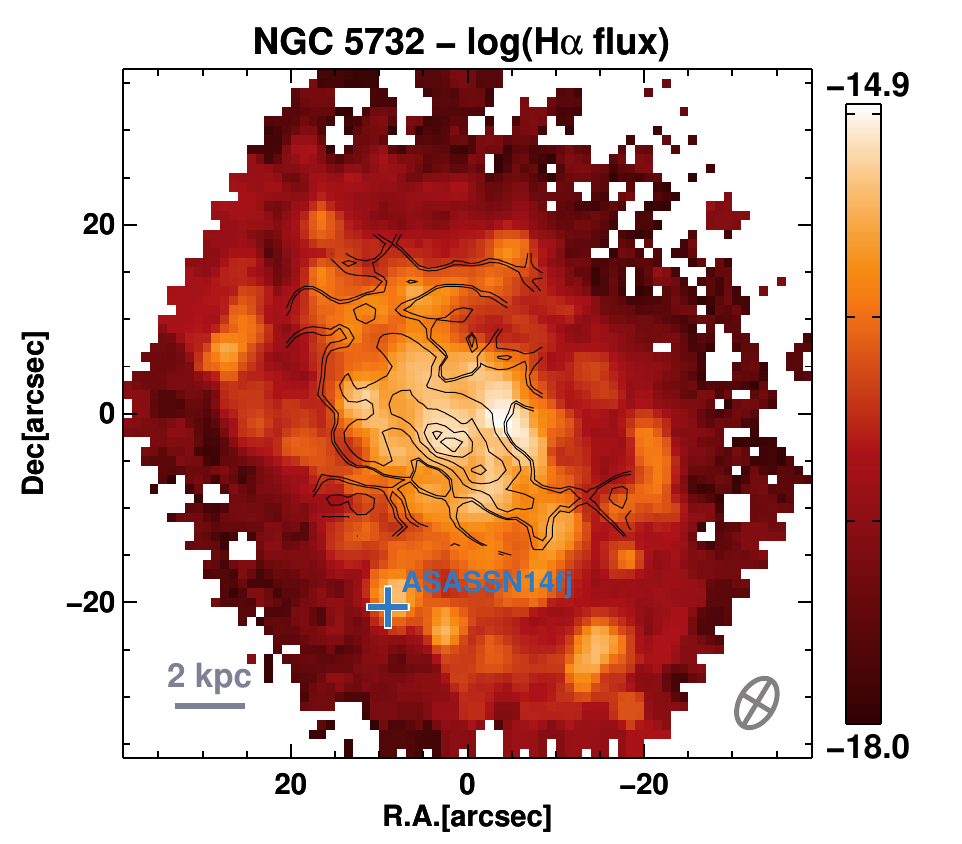}
\includegraphics[trim=0.52cm 0.1cm 0.2cm 0cm, clip=true,width=0.33\textwidth]{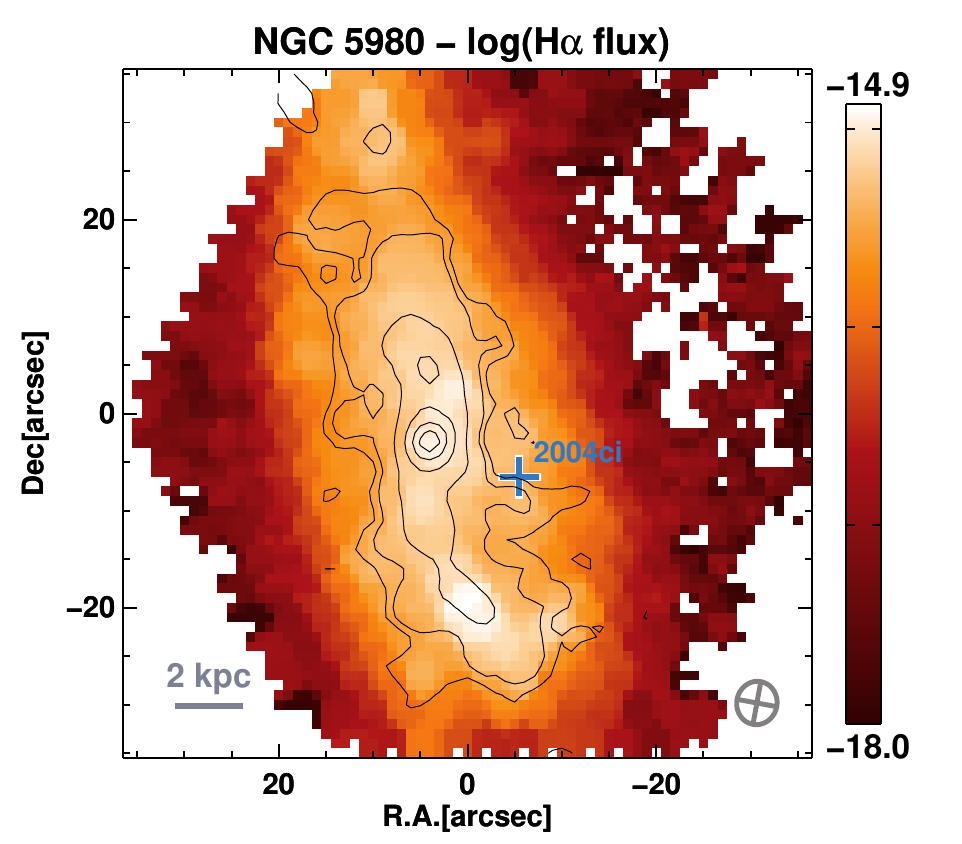}
\includegraphics[trim=0.52cm 0.1cm 0.2cm 0cm, clip=true,width=0.33\textwidth]{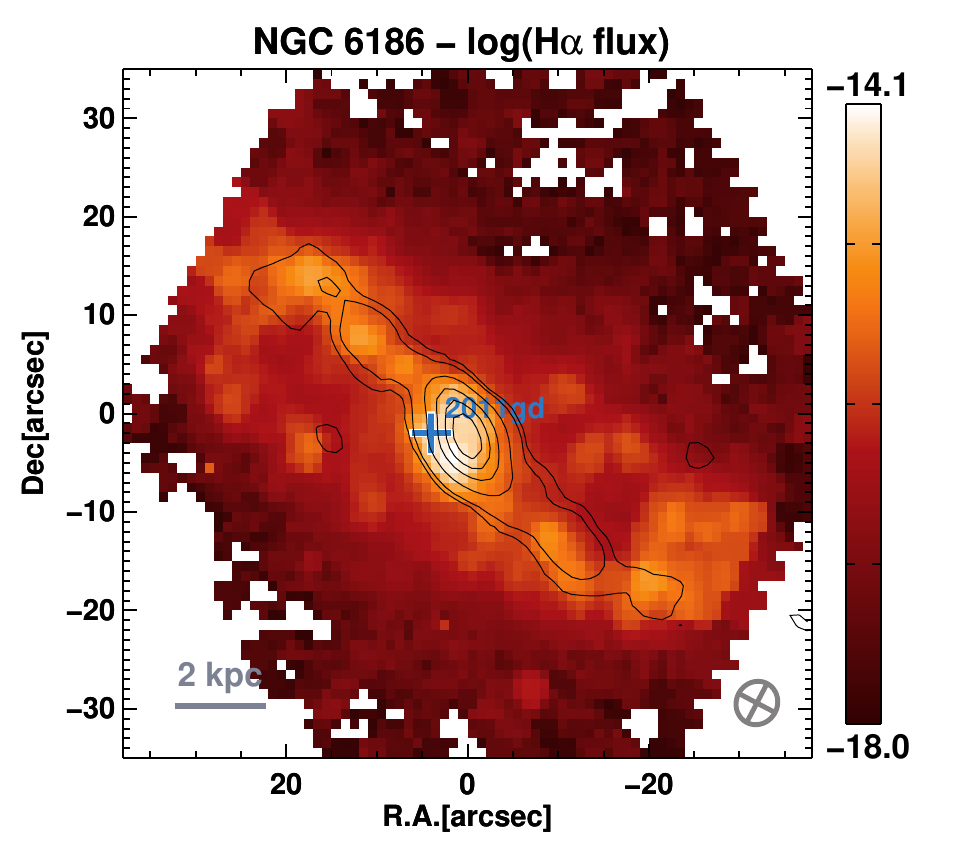}
\includegraphics[trim=0.2cm 0.1cm 0.52cm 0cm, clip=true,width=0.33\textwidth]{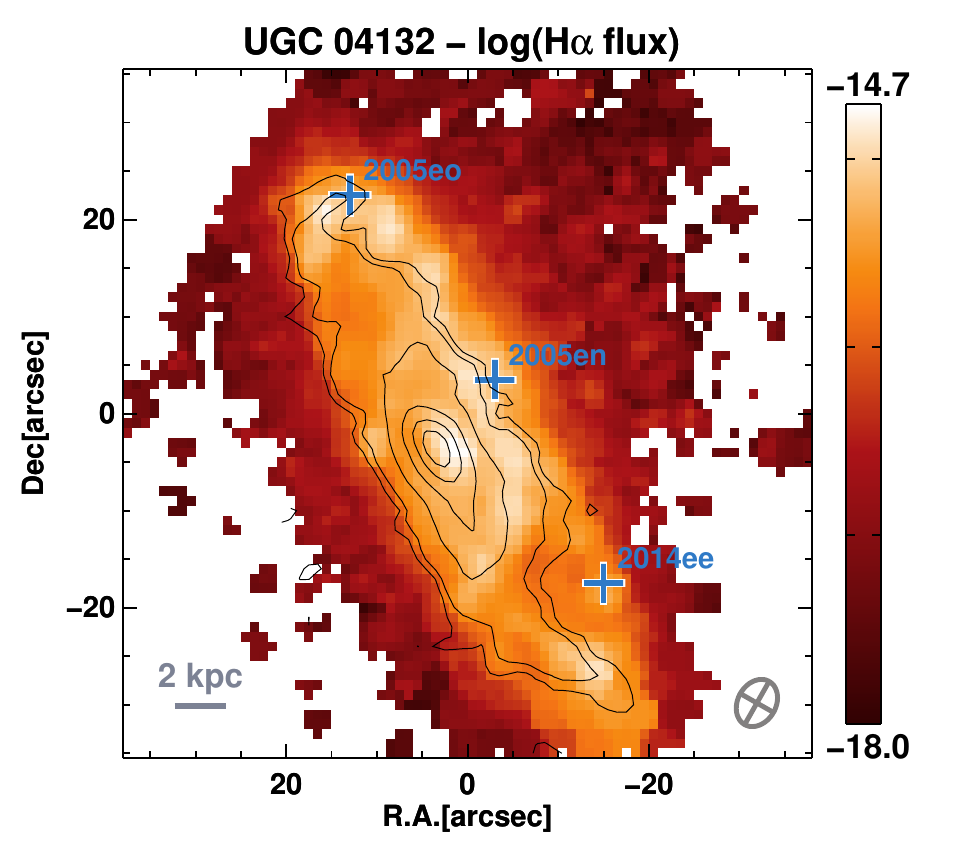}
\includegraphics[trim=0.52cm 0.1cm 0.2cm 0cm, clip=true,width=0.33\textwidth]{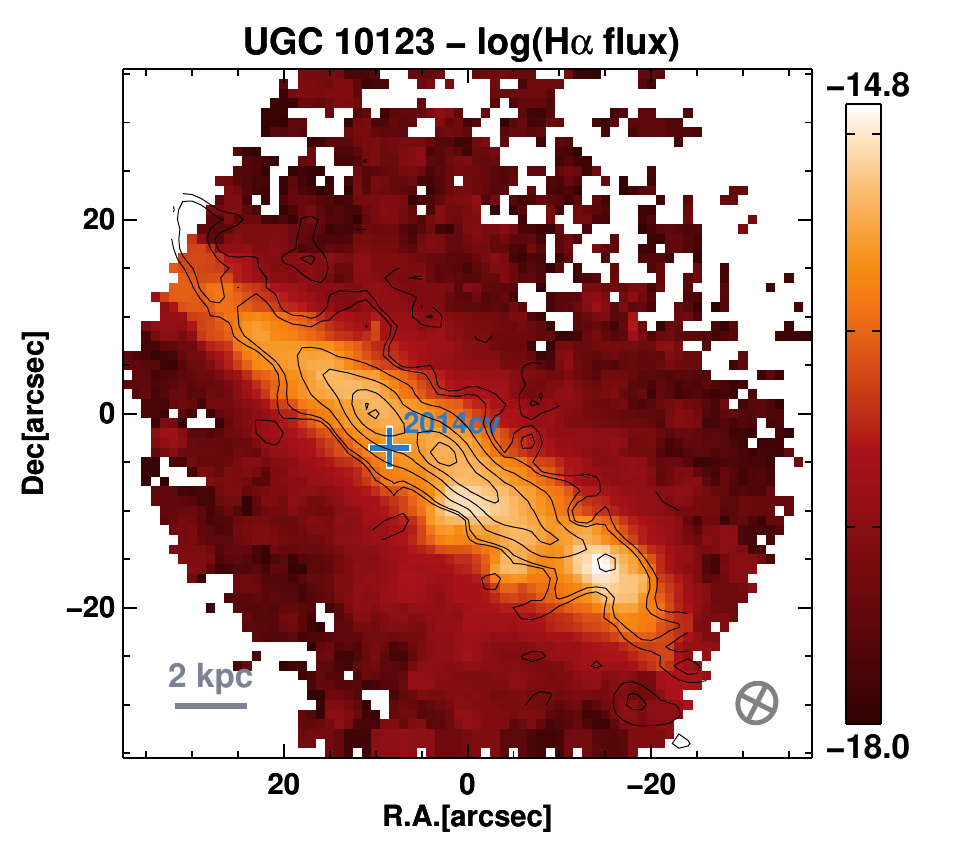}\\
\includegraphics[trim=0.2cm 0.4cm 0.52cm 0cm, clip=true,width=0.33\textwidth]{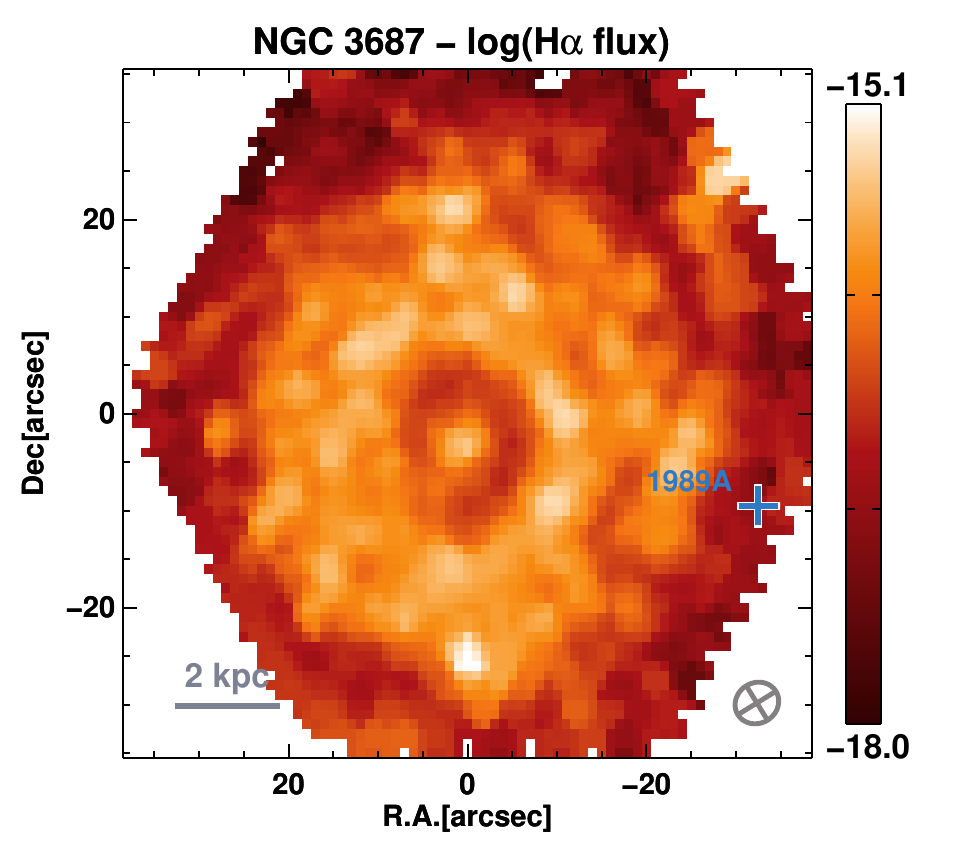}
\includegraphics[trim=0.52cm 0.4cm 0.2cm 0cm, clip=true,width=0.33\textwidth]{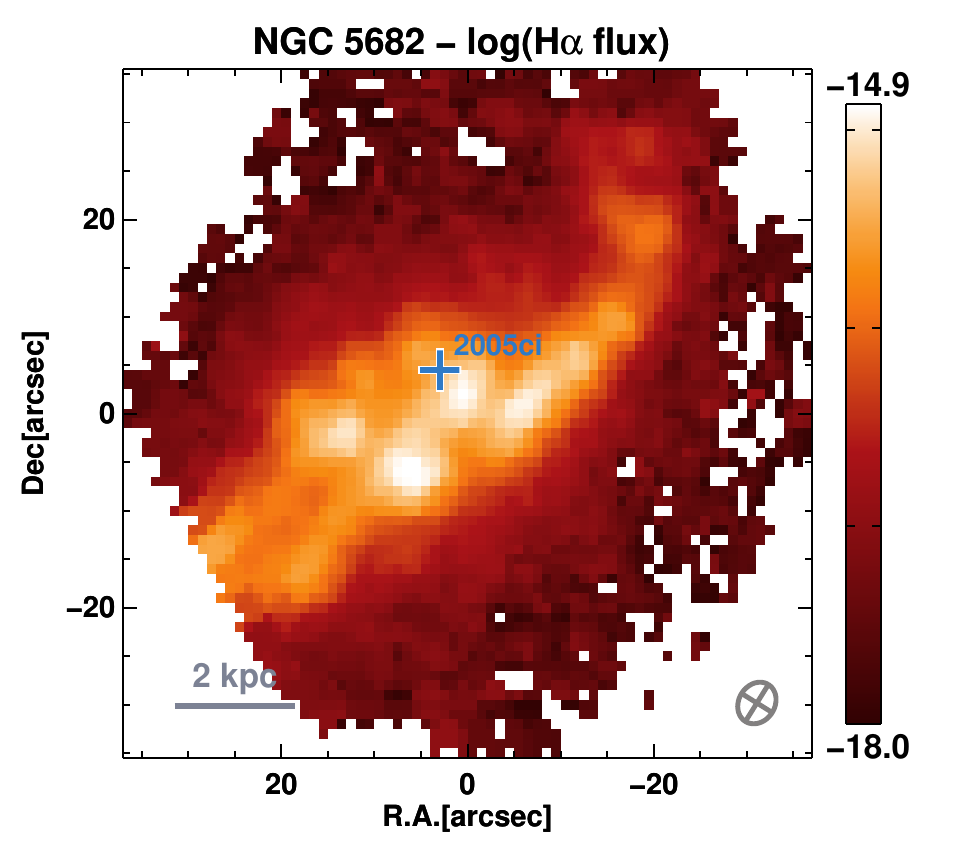}
\includegraphics[trim=0.52cm 0.4cm 0.2cm 0cm, clip=true,width=0.33\textwidth]{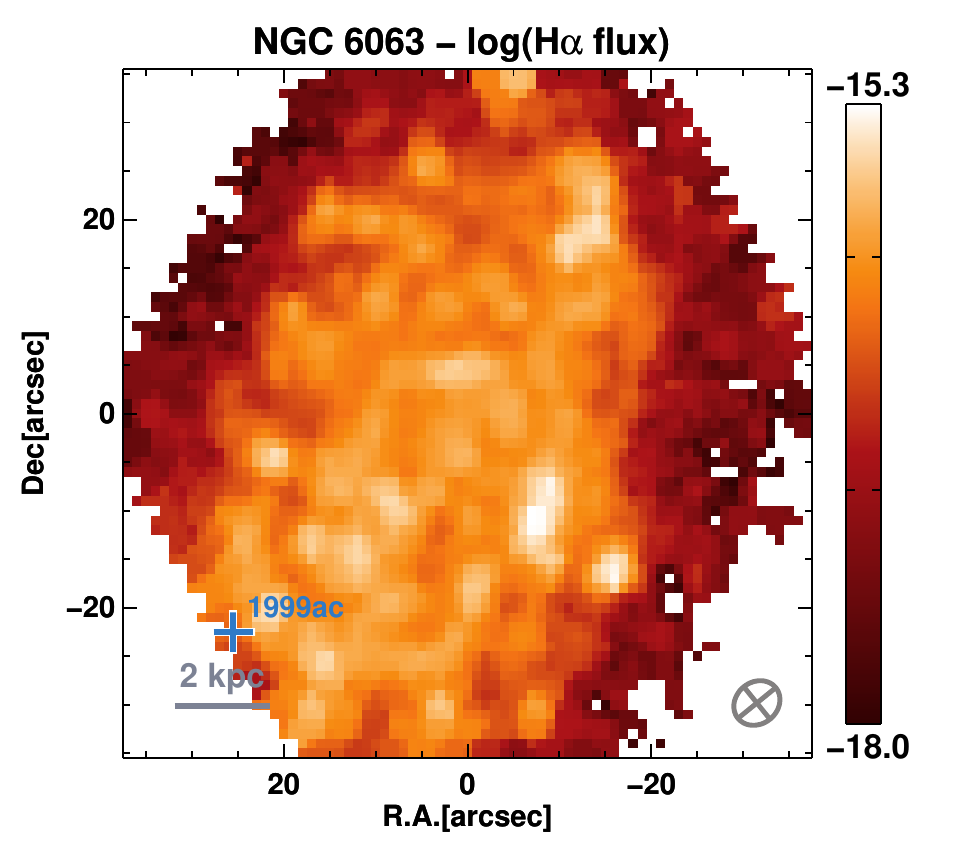}
\includegraphics[trim=0.2cm 0.1cm 0.52cm 0cm, clip=true,width=0.33\textwidth]{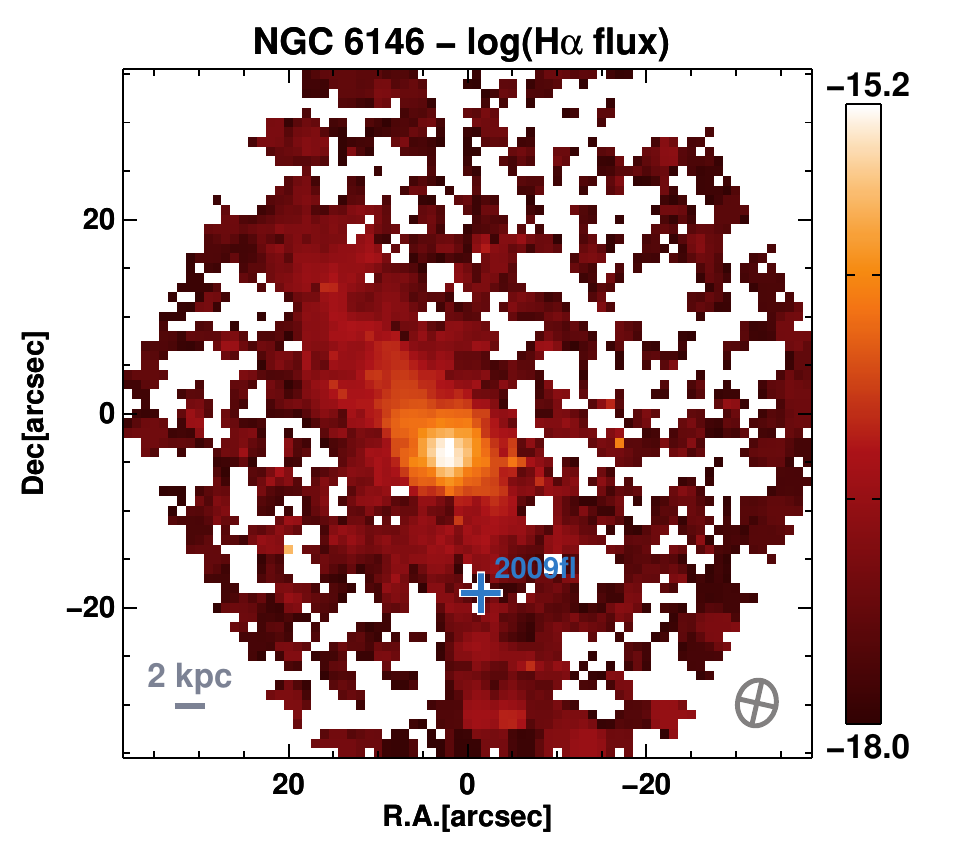}
\includegraphics[trim=0.52cm 0.1cm 0.2cm 0cm, clip=true,width=0.33\textwidth]{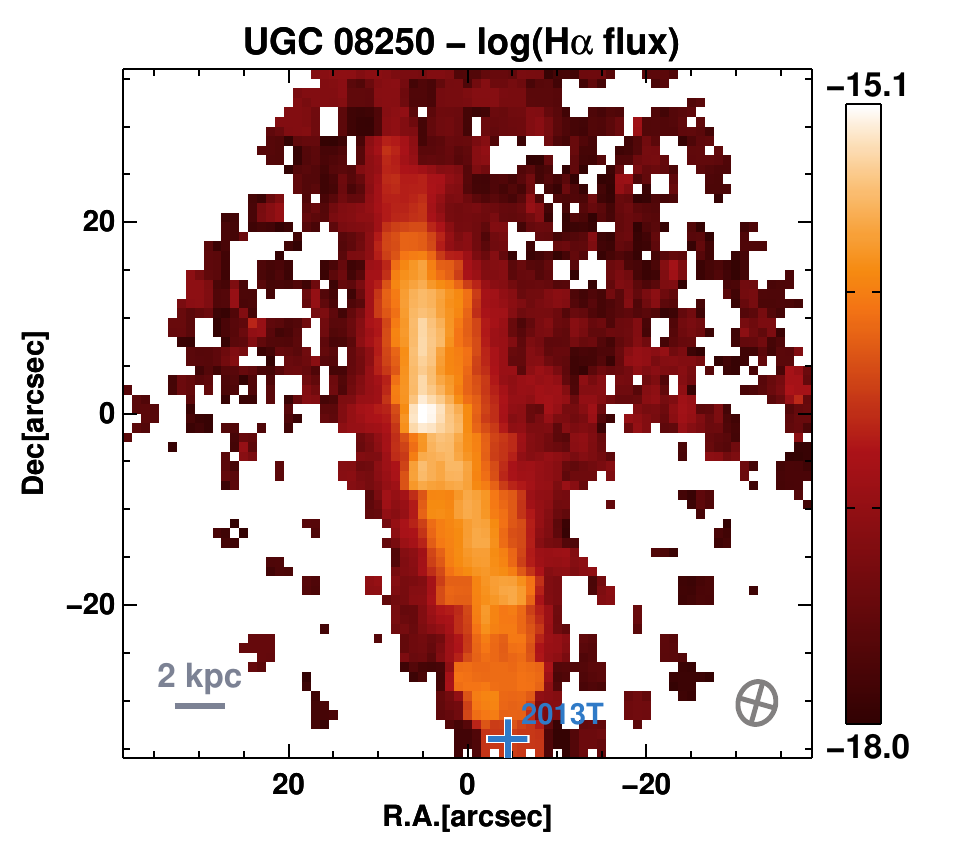}
\includegraphics[trim=0.52cm 0.1cm 0.2cm 0cm, clip=true,width=0.33\textwidth]{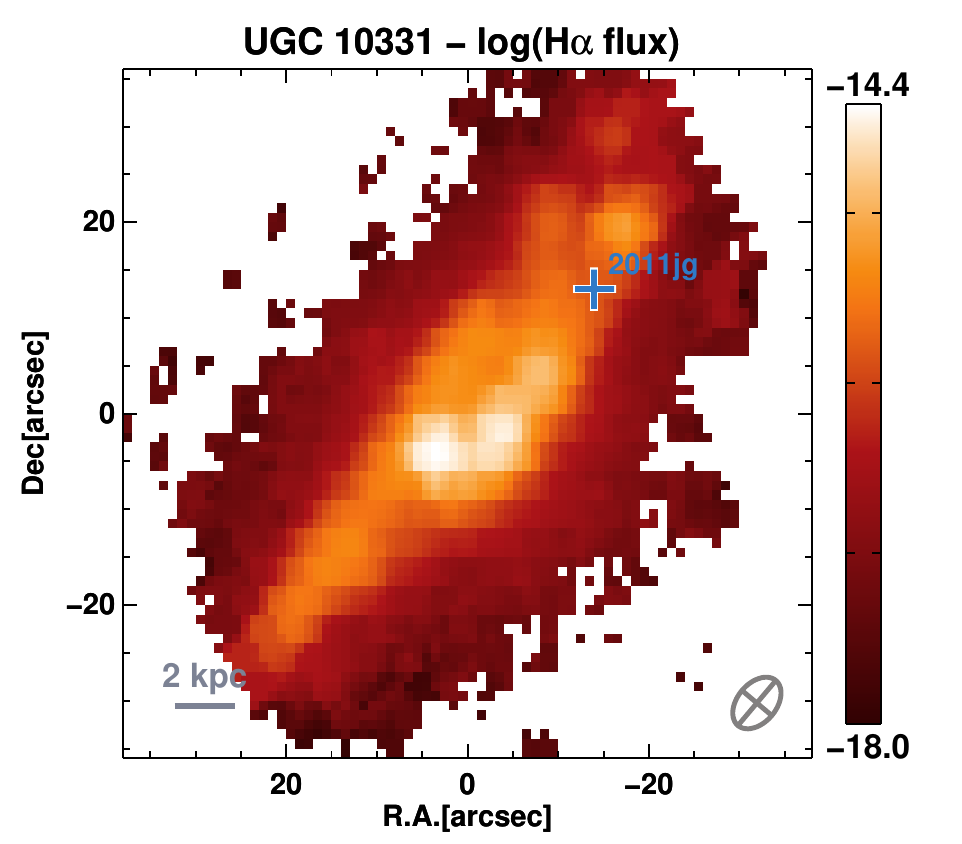}
\caption{$H\alpha$ emission maps for the 11 galaxies studied in this work with $CO$ intensities superposed in contour levels. Each contour represents the 5, 10, 30, 55, 75, and 90\% level, respectively, with respect the maximum emission. In the last two row, the 6 galaxies studied in this work with no $CO$ detection. The CARMA beam size for each map is shown in the lower right corner of each box. SN positions are marked with green crosses.}
\end{figure*}
\renewcommand{\thefigure}{\arabic{figure}}

\subsubsection{H$_2$ column density \nh} 

The molecular Hydrogen column density has been used as a proxy of interstellar matter extinction, and to correct for attenuation observations of astrophysical sources \citep{1982ApJ...262..590F}.
Being CO a proxy for \H, the CO-to-\H~conversion factor, $X_{CO}$, can be also used to estimate the column density of \H~molecular gas. For this we use the following relation,
\begin{equation}
  N (H_2) = X_{CO}~ W_{CO}, 
\end{equation}
normalized to the integration area,
where the \nh~is in cm$^{-2}$, $W_{CO}$ is the integrated line intensity over wavelength in K km s$^{-1}$ and it is referred to the ground state transition of \CO~($J = 1 \rightarrow 0$) measured in the 0th-moment map. 

We have performed these measurements using both the integrated signal of the galaxy, in order to measure the total values, and that at the SN position for the local values.

\subsection{Optical parameters from CALIFA}

Equivalent parameters of those estimated from EDGE observations in the millimeter wavelength range, were obtained from CALIFA optical data, in a similar way than in \cite{2014A&A...572A..38G,2016A&A...591A..48G}, and more details can be found there.
Here we used data reduced with the version 2.2 of the CALIFA pipeline.

In order to measure the total stellar mass (M$_*$), we used STARLIGHT \citep{2005MNRAS.358..363C} to find the combination of single stellar populations (SSPs) that best describe the observations. This was done for the integrated spectrum and for the spectrum in the spaxel at the SN position.
STARLIGHT provides the contribution of each individual SSP to the light and to the mass, where one can then obtain the star formation history (SFH), and calculate the average age and metallicity of the underlying populations, including a contribution of extinction affecting the stellar continuum (A$_V^*$).
The total stellar mass is constructed by combining the light-to-mass ratio of the different SSPs contributing to the best fit. 

Once the stellar content is well determined, it is subtracted from the observed spectrum to obtain a pure gas emission spectrum.
Each emission line is fitted using a weighted nonlinear least-squares fit with a single Gaussian plus a linear term, and converted into flux. Errors on the flux measurement were determined from the S/N of the line flux and the ratio among the fitted amplitude to the standard deviation of the underlying continuum.
For this work, we are only interested in H$\alpha$ and H$\beta$ emission lines.
While the flux of H$\alpha$ is a good proxy of the current ($<$10 Myr) star formation rate, and has been extensively used in the literature  \citep{1998ARA&A..36..189K,2015A&A...584A..87C},
the intensity ratio of the observed H$\alpha$ (6563 \AA) and H$\beta$ (4861 \AA) emission lines, i.e. the Balmer decrement, provides an estimation of the optical extinction due to the dust attenuation. 
In photoionized nebulae, the intrinsic value of the I(H$\alpha$)/I(H$\beta$) ratio is about 2.86, representative of low density nebular conditions $\sim$10$^3$ cm$^{-3}$ around a heating source with a typical temperature $T \sim 10^4$~K and large optical depths (Case B recombination; \citealt{2006agna.book.....O}).
We finally inferred \av~using the \cite{1999PASP..111...63F} MW extinction law to determine the color excess $E(B-V)$, and using the relation R$_{V}$ = \av/$E(B-V)$, assuming the Galactic average value of R$_{V}$ = 3.1.

\begin{figure*}
\centering
\begin{minipage}{0.64\textwidth}
\subfloat{\includegraphics[trim=0.9cm 0cm 0.3cm 0cm,clip=true,width=\textwidth]{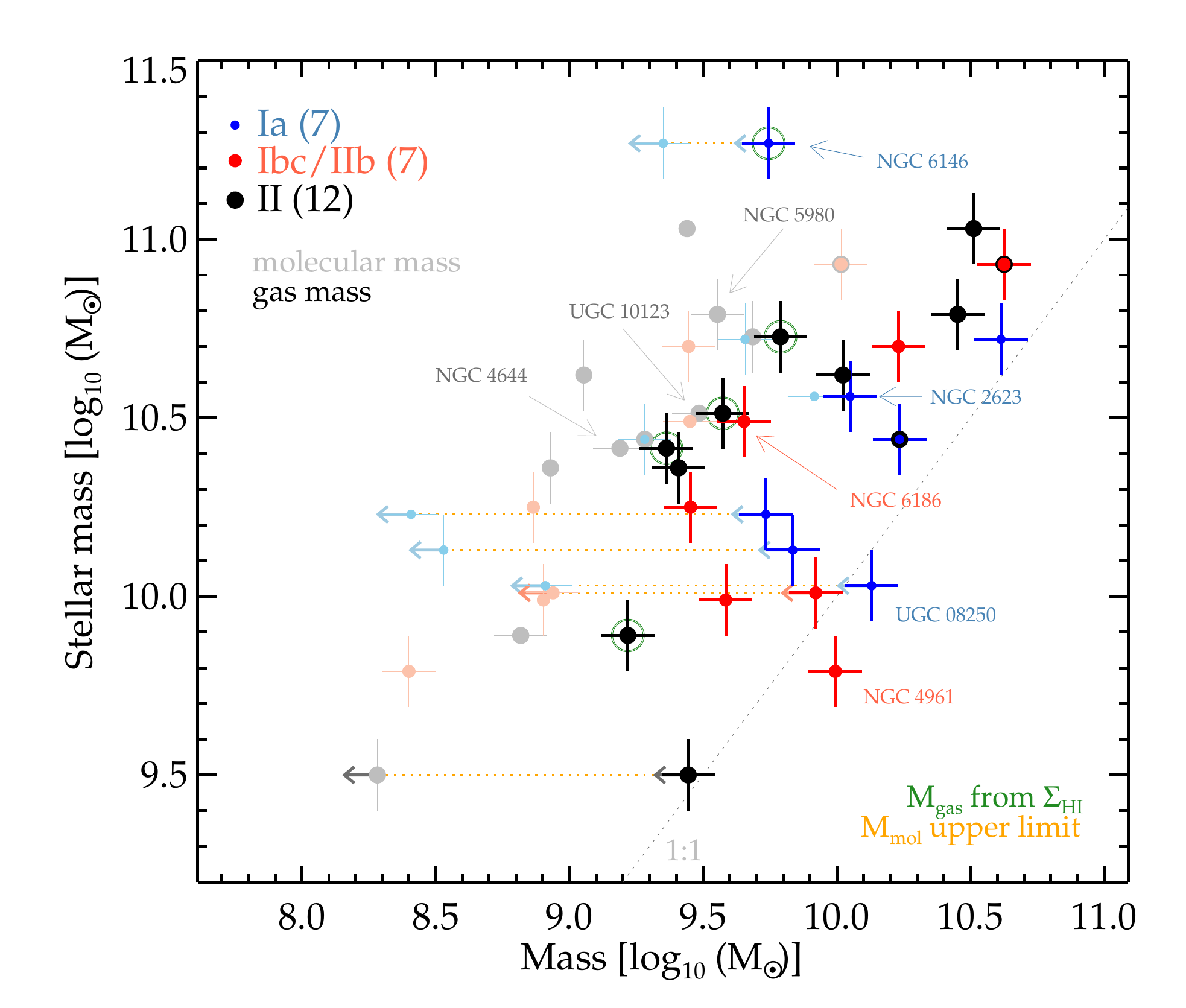}}
\end{minipage}
\qquad
\begin{minipage}{0.32\textwidth}
\vspace{6cm}
\caption{Comparison between the molecular mass ($M_{{\rm mol}}$; transparent symbols) and the total cold gas mass ($M_{gas}$; solid symbols) with respect the stellar mass ($M_{*}$), for the galaxies in our sample colored by the SN type they hosted. Dotted line represents the 1:1 relation between molecular/gas and stellar masses. 
Five galaxies are marked with green circles meaning that their HI mass has been obtained assuming an average $\Sigma_{HI}$ and therefore represent lower limits. 
Six galaxies with no CO detection in EDGE maps are plotted with left arrows representing mass upper limits.
Some measurements are accompanied with labels when the galaxy is explicitly mentioned in the main text.
}
\label{fig:mass}
\end{minipage}
\end{figure*}

In Figure \ref{fig:maps} we show  the H$\alpha$ flux maps obtained with CALIFA data for our galaxies of the sample combined with the \CO~intensity maps from CARMA in contour levels. SN position are marked with green crosses.

\subsection{Further parameters from the literature}

The atomic hydrogen H{\sc I} can be observed directly through the emission of the spin inversion line at 21~cm (1420 Mhz).
To this measurement, a factor of 1.36 factor is usually added to account for 36\% mass in Helium.
Total gas mass for the galaxies in our sample was estimated using 21~cm H{\sc I} masses reported in L\'opez-S\'anchez et al. (in prep.) and calculated with
\begin{equation}\label{totmas}
M_{\rm gas} = M_{{\rm mol}} + \left(1.36 \times M_{\rm HI}\right).
\end{equation}
Five galaxies in our sample (NGC 4644, NGC 5732, NGC 5980, UGC 10123, all hosting a SN II, and NGC 6146 hosting a SN Ia) have no atomic HI mass available in the literature.
In order to include them in our analysis we estimated the total M$_{\rm HI}$ by assuming a typical upper limit for HI mass density observed in nearby galaxies of $\Sigma_{\rm HI}$ = 10 \Msun/pc$^2$ \citep{2008AJ....136.2846B}, and multiplying this number by the area within one effective radius of each galaxy. The total gas was measured consistently using equation \ref{totmas}.

\section{Total gas mass}  \label{sec:mass}

In \cite{2014A&A...572A..38G,2016A&A...591A..48G} we found that the SN Ia host galaxy sample from CALIFA was biased to higher masses with respect the CC SN hosts, because most SNe in our sample were discovered by targeted galaxy searches, which tend to observe more massive galaxies.
In order to check how this bias affects our CALIFA/EDGE SN host galaxy sample, we first studied the mass distribution of our sample.
In this and and in following sections we also performed two-sample unweighted Kolmgorov-Smirnov (KS, \citealt{2002nrca.book.....P}) test to check whether different distributions were drawn from the same underlying population.

The average stellar mass for each of the three SN host galaxy types is around 2.5 $\times$ 10$^{10}$ \Msun~(Ia: 10.48$^{+0.24}_{-0.25}$ dex, II: 10.51$^{+0.42}_{-0.15}$ dex, Ibc: 10.31$^{+0.39}_{-0.30}$ dex), which is consistent within the uncertainties.
The lowest KS statistic is 0.36 between SN Ibc and SN II hosts (>0.83 for the other two comparison pairs), which shows that our sample is not strongly affected by this bias.
Besides the six galaxies with no CO detection, and for which we can only report upper limits, we find molecular masses between 2.51 $\times$ 10$^8$ and 1.03 $\times$ 10$^{10}$ \Msun~(Ia: 9.15$^{+0.51}_{-0.24}$ dex, II: 9.31$^{+0.37}_{-0.26}$ dex, Ibc: 9.15$^{+0.31}_{-0.24}$ dex). 
The lowest KS test is 0.65 between SNe Ia and SNe Ic.
Given the low numbers, we do not find strong evidences in favor of a heavy differences among the three SN host groups.

Combining molecular gas and stellar masses estimated in this work with atomic gas masses from L\'opez-S\'anchez et al. (in prep.), we can calculate the molecular-to-atomic gas ratio (M$_{\rm mol}$/M$_{\rm H{\sc I}}$), molecular-to-stellar mass ratio (M$_{\rm mol}$/M$_*$), and gas-to-stellar mass ratio (M$_{\rm gas}$/M$_*$) for our sample of galaxies.
In Figure \ref{fig:mass} we show the relation between M$_{\rm mol}$ and M$_{\rm gas}$ with respect to the stellar mass. All galaxies have larger stellar than gas masses, with the exception of NGC 4961 and UGC 08250 which are in turn the lightest galaxies in each SN subsample, Ibc and Ia respectively. 
This may indicate that these galaxies have still to process most of its mass into stars, while for UGC 08250 it may also be that the estimation of the M$_{\rm mol}$ may be over-predicting the amount of molecular gas.
Most of our galaxies are atomic dominated (M$_{\rm H{\sc I}}>$M$_{\rm mol}$) with a median molecular-to-atomic ratio around 0.35, but five galaxies (NGC 2623, NGC 6186, NGC 4644, NGC 5980, and UGC 10123; last three with M$_{\rm H{\sc I}}$ estimated through average density) show higher molecular mass than atomic masses. 
We find molecular-to-stellar mass ratios up to 0.12, with the exception of the merger NGC 2623 which has the highest ratio of 0.22.
For the total gas-to-stellar mass, on average, we find higher ratios ($\sim$0.29) than those found in  \cite{2015MNRAS.449.3503D} for early-type galaxies that have had a recent minor merger, which causes these objects to be as gas rich as a late-type spiral galaxies. 
In our sample we only have one early-type galaxy with no recent merger, NGC 6146. It has the largest stellar mass (11.27 dex) but does not seem to have CO present in EDGE maps (only has an upper limit for the molecular mass, 9.35 dex) and H{\sc I} mass was estimated via average density (9.39 dex). 

An important caveat on the estimation of molecular gas mass from CO emission is that a constant metallicity is assumed in the choice of X$_{CO}$. The radial gradient in metallicity implies that our choice for the CO-to-H$_2$ conversion factor could be over-predicting the molecular mass in the nucleus of these galaxies. This would increase our measurements of the molecular surface density derived from CO by some factor. 
Taking the average metallicity gradient measured in the CALIFA galaxy sample ($\sim$-0.1 dex/R$_e$; \citealt{2014A&A...563A..49S}), and the expression for the X$_{CO}$ dependence on metallicity from \cite{2012ApJ...747..124F}, we estimated that the mass at the center of the galaxy is overpredicted up to a factor 2, while we would be underestimating the mass at $\sim$2 effective radius around a factor 4. However, on average, final masses would not change much given the large uncertainties in their estimation.
Also, CO may freeze onto dust grains when temperatures drop below 20 K, underestimating the reported masses. A major assumption which may affect our extracted masses is the optical thinness of the gas. This is a safe assumption for low density environments of $<10^2$~cm$^{-3}$. For denser regions, $^{13}$CO would be a better mass tracer.

From the results of this section we conclude that, although SNe Ia usually explode in more massive galaxies (they do in ellipticals while CC SNe do not), our SN host galaxy sample show similar properties in terms of mass, and is not biased towards any particular direction.
There is only one elliptical galaxy in our SN Ia sample, so we are mainly probing the properties of SN Ia late-type hosts.
The fact that no significant differences are found, stresses that any local differences shown in the following sections does not come for differences in the total properties of galaxies, with the only caution of the small sample size.

\section{SFR and SN type} \label{sec:sfr}

Since stars form in molecular clouds, it would be reasonable to expect correlations between CO and tracers of star formation such as the H$\alpha$ emission from ionized gas, which is associated with massive young stars.

\begin{figure*}
\centering
\includegraphics[trim=0cm 0cm 0cm 0cm, clip=true,width=\textwidth]{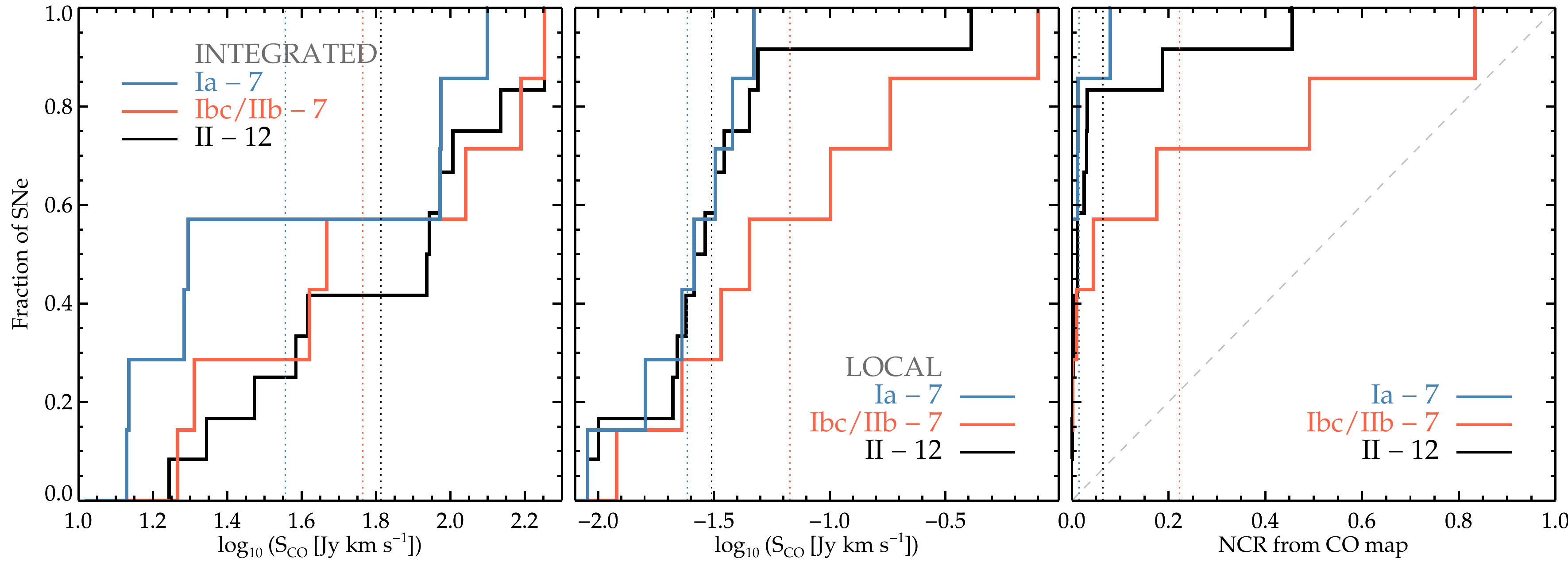}
\includegraphics[trim=0cm 0cm 0cm 0cm, clip=true,width=\textwidth]{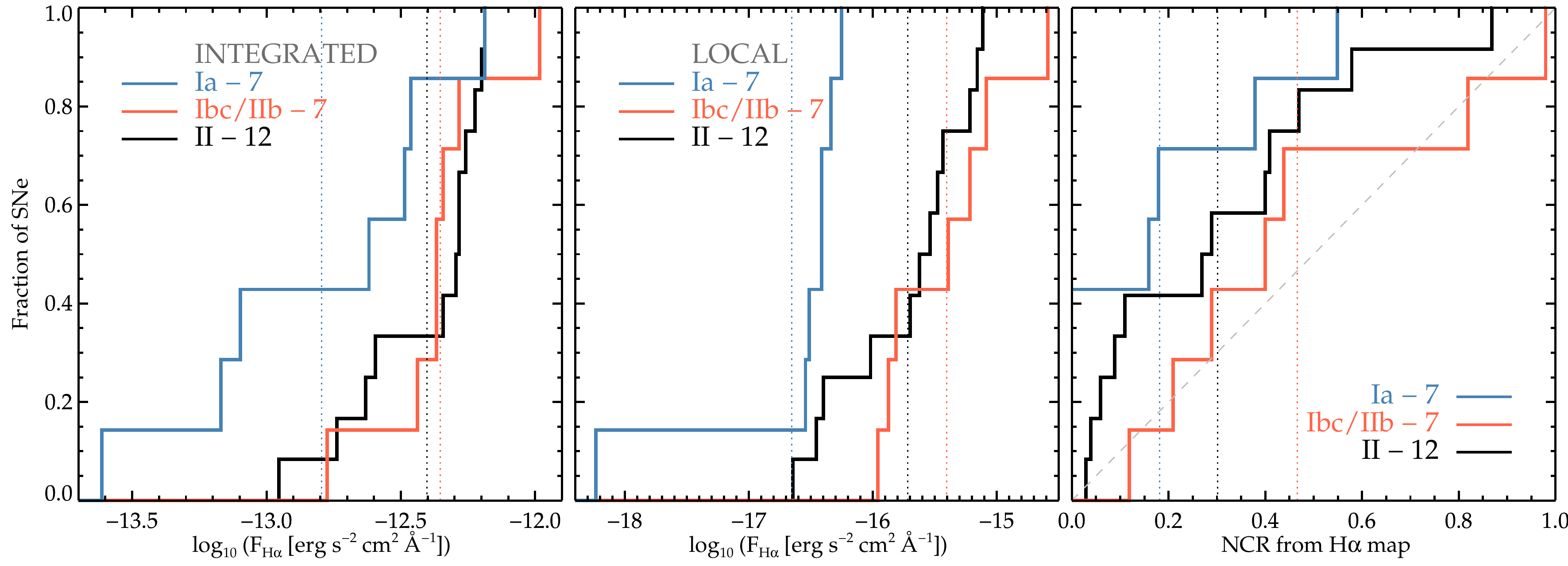}
\caption{Cumulative distributions of the CO intensity (top row) and the H$\alpha$ flux (bottom row) measurements integrating all the emission of the galaxy ({\it total}, left), at the SN locations ({\it local}, middle), and the NCR pixel statistics ({\it right}), for the three SN types. Vertical dotted lines represent the averages of the distributions.}
\label{fig:sfr}
\end{figure*}

Figure \ref{fig:sfr} shows the distributions of the CO intensity measured both from integrating the signal of the whole extent of galaxies, and at the SN location, for the three SN types. As a comparison, we show similar distributions of the H$\alpha$ flux of the total galaxy and at the local SN environment.
Left panels in Figure \ref{fig:sfr} show the total/integrated S$_{CO}$ and F(H$\alpha$) distributions. While the CO intensity of the three groups are quite similar (lowest KS$\sim$0.17 for SNe II-Ia), the SN Ia total H$\alpha$ flux shows lower values than both CC SN subsamples.
This difference is only statistically significant (although we note the low numbers) for SN Ibc and SN Ia hosts (KS=0.02).
While CCSNe are only found in galaxies with ongoing and strong star formation, SNe Ia can be found in any kind of galaxy, including elliptical and passive galaxies. Moreover, their progenitors are older stars that were born at places not necessarily connected with their explosion locations given their longer lifetimes.
In our sample there is only 1 elliptical galaxy (NGC 6146), which shows the lowest H$\alpha$ flux as expected.

Regarding the local distributions (see middle top and bottom panels in Fig. \ref{fig:sfr}), we find that the sequence from higher to lower CO intensity is ordered similarly than the optical distributions. SNe Ibc explode at locations with higher current star formation rates but also that contain a higher amount of molecular gas, probably indicating that star formation will be occurring at that position for more time.
This is in agreement with the H$\alpha$ results, where the SN Ibc distribution is shifted to higher F(H$\alpha$) values.
The SN Ia distribution is located in both cases, F(H$\alpha$) and CO, at the lower values, indicating that they have a worse relation to star formation. 
SNe II are located in the middle of the two other distributions in both CO and H$\alpha$ indicators. 
However, while in the optical the SN II distribution is not significantly different from the SN Ibc distribution (KS$_{II-Ibc}$=0.61, KS$_{II-Ia}$=0.01, KS$_{Ibc-Ia}$=0.01), in the millimeter the SN II distribution is quite similar to the SN Ia distribution (KS=0.74) and different from SN Ibc distribution (KS$_{II-Ibc}$=0.21, KS$_{Ibc-Ia}$=0.12).
These distributions show that SNe Ibc occur at places where SF will continue occurring while SNe II tend to occur at locations with less gas reservoir.
This result is reinforced by the fact that around seventy percent of SNe Ia (6 over 7, 86\%) and SNe II (6 over 12, 50\%) in our sample exploded at locations where there was no signs of CO, or at least it was below the detection limit of CARMA. Upper limits were reported at these positions, and the difference to SNe Ibc would be in any case larger.

\subsection{Spatially resolved statistics}

\cite{2012MNRAS.424.1372A} presented a statistical method, which they named normalized cumulative rank (NCR) pixel function, to study the correlation of different SN types to the local H$\alpha$ intensity used as a proxy to the very recent ($\lesssim$10 Myr) star formation rate. 
The construction of the NCR function in a given galaxy basically consists of sorting the H$\alpha$ flux values in increasing order, form the cumulative distribution, and normalize this to the total emission of the galaxy. This associates each pixel with an NCR value between 0 and 1, where 1 is the brightest pixel and 0 all pixels without emission. 
Assuming that the H$\alpha$ emission scales by the number of stars that are formed \citep{1994ApJ...435...22K}, a flat NCR distribution  (or diagonal cumulative NCR distribution) with a mean value of 0.5 would mean that this type of SN accurately follows the stars that are formed and mapped by that particular SF tracer.
The same method was also used by \cite{2013MNRAS.436.3464K} with UV imaging, which traces populations of around populations $\sim$10-100 Myr.
While \cite{2012MNRAS.424.1372A} found that SNe Ic was the SN type that best followed the NCR-H$\alpha$ diagonal, thus concluding that their progenitor stars have ages of around 10 Myr, \cite{2013MNRAS.436.3464K} found that SNe II was the type following better the NCR-UV diagonal distribution.

We computed the NCR cumulative distributions in both our CO and H$\alpha$ intensity 2D maps. Right panels in Figure~\ref{fig:sfr} show the resulting NCR distributions for each sub-sample.
For the NCR-H$\alpha$ distribution we recover previous results: SNe Ibc have higher average values ($\langle NCR \rangle$=0.46) and also higher KS statistic when compared with the diagonal distribution (KS=0.88); SNe II follows with $\langle NCR \rangle$=0.30, and finally SNe Ia with $\langle NCR \rangle$=0.18, and both with KS statistics with respect to the diagonal lower than 0.4.
Given the lower numbers of our sample we find relatively large KS test numbers, but the Ibc-Ia is 0.13, and the other two combinations are $\sim$0.3.

Regarding the NCR-CO distributions, the SN~Ibc sample is the closest to a flat distribution ($\langle NCR \rangle$=0.23) thus closely tracing the cold gas reservoir in the galaxy. 
Both SNe II and SNe Ia show NCR-CO low values, 0.06 and 0.02 respectively. 
The two-sample KS tests between the NCR distributions indicates that SN~Ia and SN~II samples come from similar underlying distributions with a probability of 0.6. 
The SN~Ibc distribution is the most different from the other two, however, the probability that it came from the same distribution as SNe~II is 0.4. 
The KS tests between each distribution and the diagonal distribution support the fact that SNe Ibc (0.42) is more similar to the diagonal than the SN II (4.1e-3) and SN Ia (9.1e-04) distributions.
Again, although all these results should be taken with caution, our results show that SNe Ibc have higher NCR values when studying CO and that the separation between SNe Ibc and SNe II is larger in CO than in H$\alpha$ (KS$_{NCR-CO}$=0.41, KS$_{NCR-H\alpha}$=0.33).

\section{N(\H) and A$_V$ extinction estimates}  \label{sec:ext}

\begin{figure}
\centering
\includegraphics[trim=1cm 0cm 0.9cm 0.9cm, clip=true,width=\columnwidth]{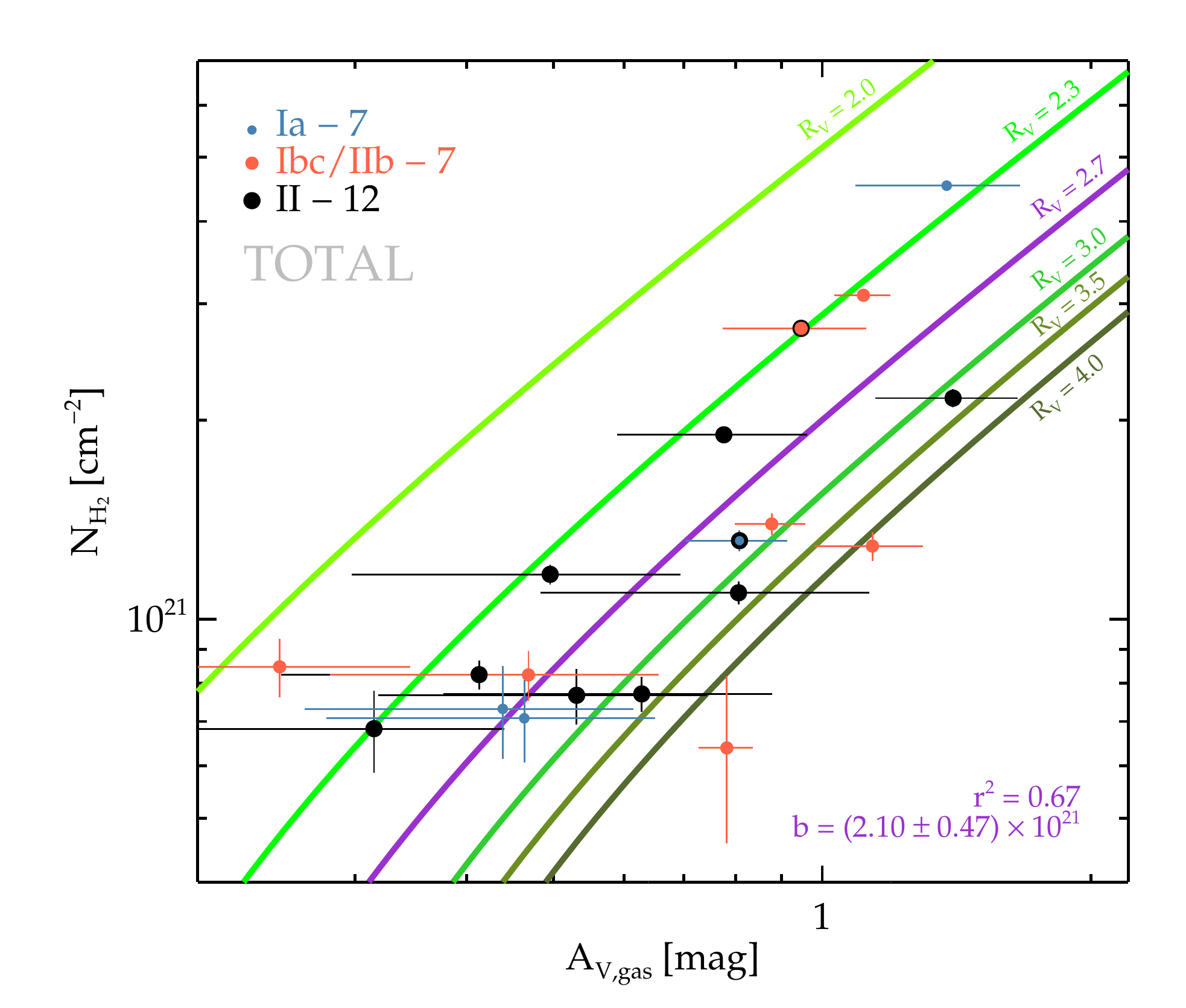}
\caption{Molecular column density, \nh, with respect the optical extinction derived from the Balmer decrement, A$_{V}$, for the integrated measurements of our galaxy sample. The slope of the linear fit (in purple) to the data gives the relation between the two parameters. The corresponding slopes for different values of R$_V$ are shown. Our data shows a preference for R$_V$ = 2.7, a value slightly lower than the average MW standard.}
\label{fig:nh2rel}
\end{figure}

\begin{figure}
\centering
\includegraphics[trim=1.0cm 0cm 0.9cm 0.9cm, clip=true,width=0.92\columnwidth]{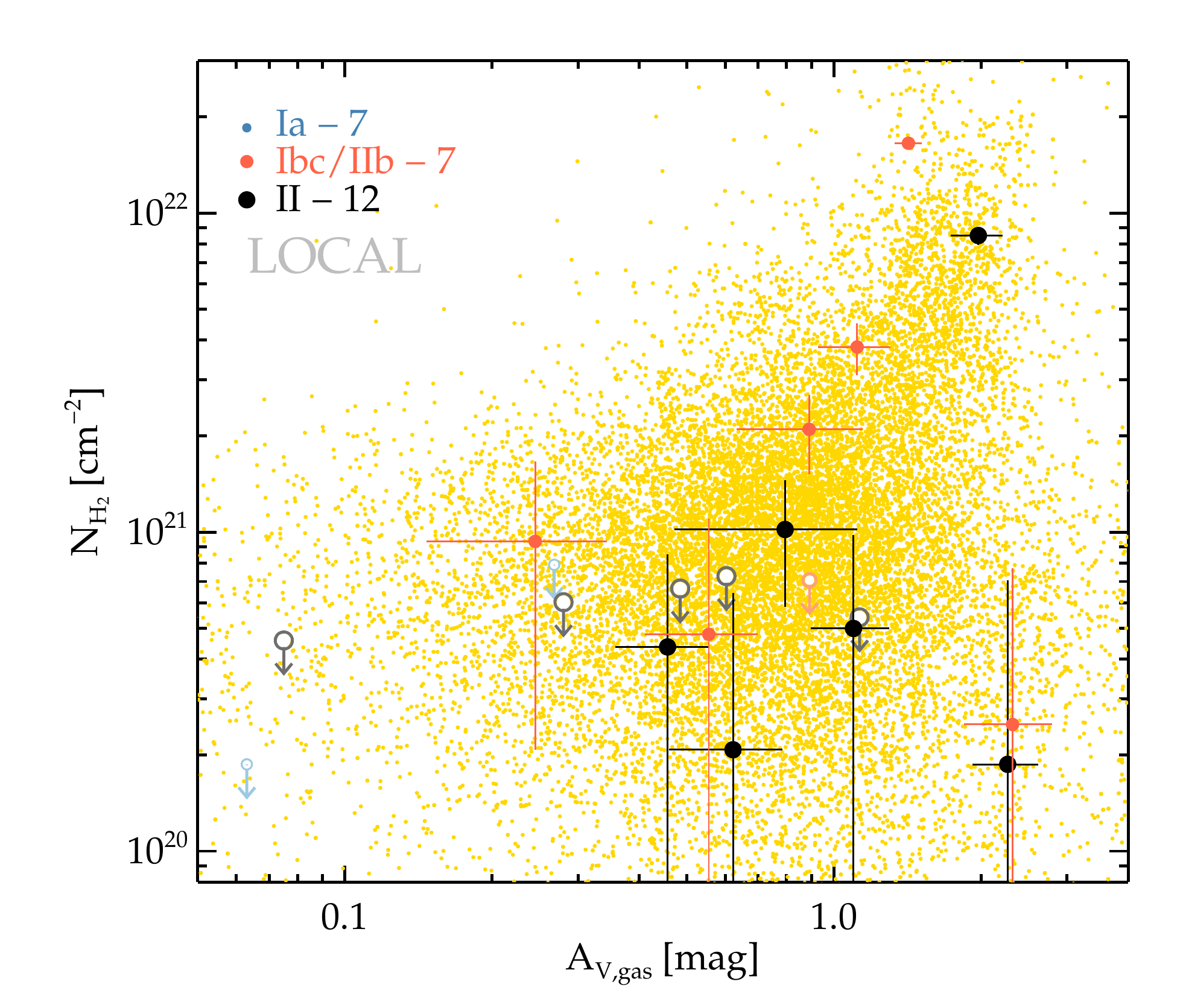}
\includegraphics[trim=1.0cm 0cm 0.9cm 0.9cm, clip=true,width=0.92\columnwidth]{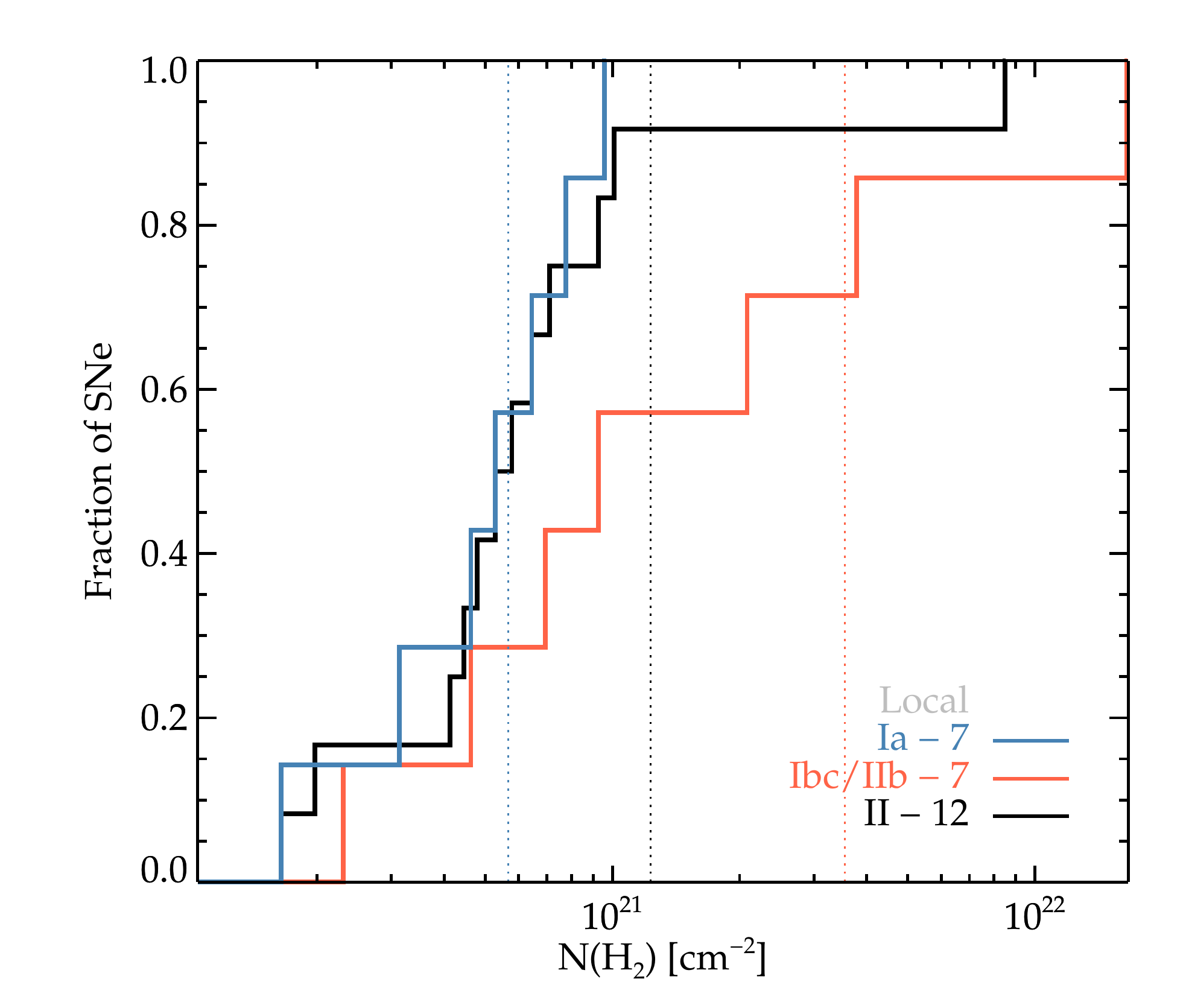}
\includegraphics[trim=1.0cm 0cm 0.9cm 0.9cm, clip=true,width=0.92\columnwidth]{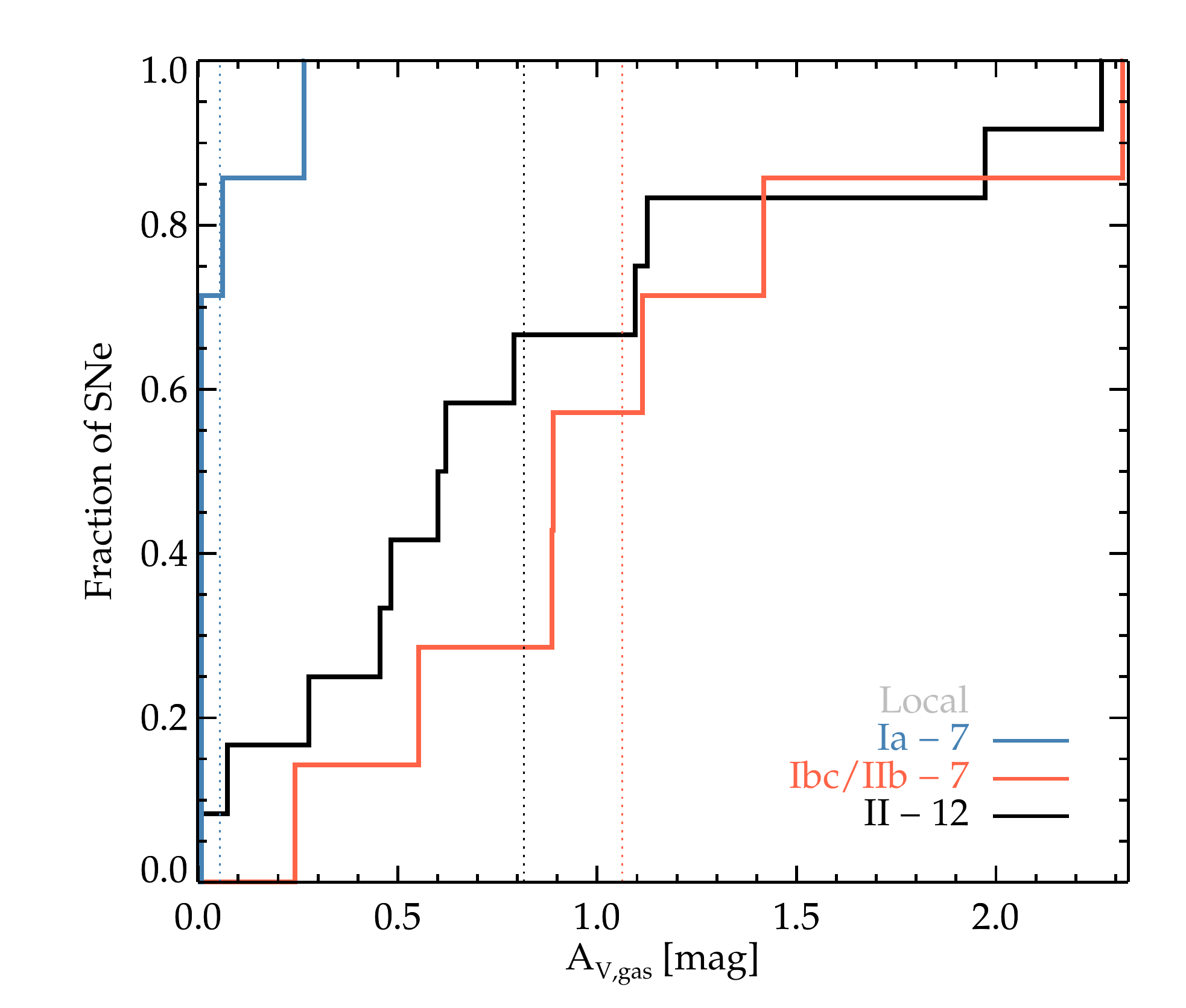}
\caption{Top panel: Molecular column density, \nh, with respect the optical extinction derived from the H Balmer decrement, A$_{V}$, for the measurements at the SN positions, where solid points indicate actual measurements and empty points represent upper limits.
In the background a cloud of measurements for all individual positions in all 23 galaxies.
Mid and bottom panels: Distributions of column density ($N_{H_2}$, middle) and local visual extinction ($A_V$, bottom) for the three SN types. For most SNe Ia these are upper limits for both $A_V$ and $N_{H_2}$, and for most SNe II these are $N_{H_2}$ upper limits.}
\label{fig:locext}
\end{figure}

\begin{table*}\footnotesize
\caption{References for SN photometry and spectroscopy used in Section \ref{sec:sn}.}
\label{tab:ref}
\centering
  \begin{tabular}{lccc}
    \hline\hline
    SN name & Type & Photometry & Spectroscopy  \\
    \hline
   2001en &  Ia & \cite{2010ApJS..190..418G,2009ApJ...700..331H} & \cite{2012AJ....143..126B,2012MNRAS.425.1789S}  \\ 
   1999gd & Ia &  \cite{2006AJ....131..527J} &   \cite{2012AJ....143..126B,2012MNRAS.425.1789S} \\
   1989A  & Ia &  \cite{1991AA...247..410B,1990AA...236..133T} & --- \\
   1969C & Ia &  \cite{1990AAS...82..145C} & --- \\
   1999ac & Ia-pec & \cite{2006AJ....131.2615P,2006AJ....131..527J} &  \cite{2012AJ....143..126B,2012MNRAS.425.1789S} \\
   2005en & II & --- & CfA SN archive \\
   2014ee & IIn & --- &  \cite{2014AN....335..841T} \\
   2005ip & IIn & \cite{2012ApJ...756..173S} & CSP, (Guti\'errez et al. in prep.) \\
   2005au & II & \cite{2012ApJ...756L..30A} & CfA SN archive \\
   2005ci & II & \cite{2016AA...588A...5T} & \cite{2016AA...588A...5T}; CfA SN archive \\
   2004ci & II & --- & CfA SN archive \\
   2005eo & Ic &  \cite{2007PhDT.........7M,2011ApJ...741...97D} & \cite{2014AJ....147...99M} \\
   2005az & Ic & \cite{2011ApJ...741...97D} & \cite{2014AJ....147...99M} \\
   2003el & Ic & --- & CfA SN archive \\
   1988L & Ib & --- & \cite{2001AJ....121.1648M,2001AAS...199.8408R}\\
    \hline
\end{tabular}

$*$ All CfA SN archive data downloaded via the Open Supernova Catalogue \citep{2016arXiv160501054G}
\end{table*}

The amount of visual extinction along a typical line-of-sight through the ISM is strongly correlated with the total column density of molecular Hydrogen. 
In Figure \ref{fig:nh2rel} we show the relation between the estimations of both the integrated molecular Hydrogen column density, \nh, with respect to the optical extinction derived from the H Balmer decrement, \av.
The three galaxies with upper limits on the CO detection, are far below the region shown in the Figure, and are not shown for the sake of clarity.
Our measurements are best fit with a first order polynomial of the form,
\begin{equation}
\frac{N(H_2)}{\rm cm^{-2}} = (-0.20 \pm 0.38) 10^{21} +  (2.10 \pm 0.47) 10^{21}  \frac{A_V}{\rm mag}.
\end{equation}
This is consistent with previous reported results in the literature: 
\cite{1978ApJ...224..132B} derived the current 'standard' conversion between total visual extinction, \av, and total Hydrogen column density, \nh, in the diffuse ISM, 1.87 $\times$10$^{21}$ cm$^{-2}$ mag$^{-1}$,  to be representative of dust in diffuse regions.
\cite{2009MNRAS.400.2050G} measured the same relation using X-ray observations of 22 SN remnants, and reported a slope of (2.21 $\pm$ 0.09) 10$^{21}$ [cm$^{-2}$ mag$^{-1}$]. 
Previous work by \cite{1973A&A....26..257R}, \cite{1975ApJ...198...95G}, and \cite{1995A&A...293..889P} found similar values (1.85, 2.22$\pm$0.14, 1.79$\pm$0.03, respectively).
Although our result is consistent but slightly larger than those reported in these works, we note that while all these previous estimations were local (z$\approx$0), the average redshift of our galaxy sample is 0.0137(5). 
Also, \cite{2002ApJ...577..221R} found that this factor is sensible to R$_V$, and \cite{2003ARA&A..41..241D} showed that it can be higher for lower R$_V$ values. 
Reordering his equation 5, and assuming \cite{1999PASP..111...63F} reddening law and A$_I$/\av~ = 0.554 the expression between \nh~and \av~can be expressed as,
\begin{equation}
\frac{N(H_2)}{\rm cm^{-2}} = 0.554 /  \left[0.296 - 0.355 \left(3.1/R_V - 1\right)\right] 10^{21}  \frac{A_V}{\rm mag}.
\end{equation}
The corresponding slopes for different values of R$_V$ are shown in Figure \ref{fig:nh2rel}.
In particular, the relation found with our data correspond to an $R_V$ of about 2.7, which is slightly lower than the standard 3.1, and qualitatively in agreement with other R$_V$ estimations made directly from SN Ia \citep{2007ApJ...659..122J,2010ApJ...722..566L,2010AJ....139..120F} and SN II \citep{2010ApJ...715..833O,2014AJ....148..107R,deJaeger2017} photometry.

Regarding the local values at SN positions, we present the distributions of $A_V$ and $N({H2})$, and their relation in Figure \ref{fig:locext}. 
In the top panel we added all individual measurements at each spaxel in the 23 galaxies. Filled symbols indicate actual measurements, and empty circles are upper limits. 
A dense cloud of points can be recognized at \av $\sim$0.8 mag and \nh$\sim$7$\times$10$^{20}$ cm$^{-2}$, with an arm towards higher values in both parameters.
In the bottom panels, we show cumulative distributions of the measurements at SN locations.
The $N_{H2}$ distribution is similar to the S$_{CO}$ local distributions, since their measurement only differ by a factor X$_{CO}$.
We see again that SNe Ibc occur at places where the H$_2$ is denser than at the locations of other SN types. 
\av~ is not directly dependent on F(H$\alpha$), since it includes the ratio of this flux to F(H$\beta$). However we find that SNe Ia in our sample do not seem to be heavily affected by dust extinction, while both SNe II and SNe Ibc are very reddened.
SNe Ibc tend to occur at places affected by both higher column densities and large amount of dust. On the other hand SN Ia locations are usually less affected by extinction in the optical and have lower H$_2$ column densities. Interestingly, for SNe II we find low H$_2$ column densities and larger amounts of dust.
We note that although the distribution of local N(H$_2$) at SN Ia positions may look similar to the SN II distribution, this is only an artifact of the upper limits on the CO detection. While only 1 over 7 (14\%) SNe Ia have actual CO detections, we found CO at 6 over 12 (50\%) SN II locations. This may shift the SN Ia distribution to even lower values.  

\section{Supernova extinction estimates} \label{sec:sn}

Extinction towards SN light has been extensively studied in recent years.
The most intriguing result is that the general extinction law, expressed in the total-to-selective extinction ratio, $R_V$, of SNe seems lower than the average in the Milky Way ($R_V^{MW}=3.1$; e.g. \citealt{2007ApJ...664L..13C,2009ApJ...700.1097H,2010ApJ...715..833O,2014AJ....148..107R}) , although this effect could be caused by improper color models \citep{2014ApJ...780...37S,2011A&A...529L...4C}.  Regardless, the most reddened SNe Ia consistently show very low $R_V$ values \citep{ 2014ApJ...788L..21A,2016A&A...590A...5G} posing a real challenge to understand extinction towards these SNe. Recent observational evidence suggests that this may also be caused by the presence of circumstellar material (CSM) around SN progenitors, which would also affect SN colors \citep{2008ApJ...686L.103G,2013ApJ...772...19F}. 
However, other studies suggest that the dust responsible for this observed reddening is predominantly located in the interstellar medium (ISM) of their host galaxies (e.g \citealt{2013ApJ...779...38P}), which would imply different nature of dust compared to the average vicinity of the Milky Way.
Determining the contribution of CSM in the total SN reddening would have severe implications in SN science. 
It would provide definitive information on the SN Ia progenitor scenario in play. For instance, in the single degenerate scenarios it would not be unreasonable to expect H and/or He rich gas from the companion star to be present in the CSM \citep{2012Sci...337..942D}, although no such emission has been seen in normal SNe Ia \citep{2013MNRAS.435..329L}.
For SNe II, low R$_V$ values have also been reported, and it has also been proposed that CSM of different sizes and densities may explain the shorter rise times and early spectral features of some objects \citep{2015MNRAS.451.2212G, 2016ApJ...818....3K}.

With our environmental data in hand, we hope to bring new insight into the nature of the higher extinction and peculiar reddening law found at some SN locations, to understand its origin.
We have compiled photometric and spectroscopic data of a number of SNe that were available from different sources, and used different approaches to get alternative estimations to the extinction towards the SN. 
All spectra and photometry have been corrected for MW extinction using a \cite{1999PASP..111...63F} law and the dust maps of \cite{2011ApJ...737..103S}. 
All references to the sources can be found in Table \ref{tab:ref}.

\subsection{Spectroscopic method: Na I D absorption}

The blended Na {\sc I} D $\lambda\lambda$5890, 5896 narrow absorption pseudo-equivalent width has extensively been used in several works as a proxy for interstellar extinction within the Milky Way \citep{1974ApJ...191..381H}, as well as in extragalactic sources \citep{2003fthp.conf..200T}.
The strength of the line can be a useful diagnostic of the amount of absorption by intervening material in the line of sight. 
\cite{2012MNRAS.426.1465P} derived a relation between the Na I D pseudo-equivalent width (pW) and the color excess, E(B-V), which assuming an extinction law and a value for R$_V$ permits an estimation of \av.
However, its reliability has also been put in question.
\cite{2013ApJ...779...38P} showed that, although this relation follows well Na I D absorptions produced by Milky Way extinction, there was a relatively large fraction of SNe Ia with strong Na I D absorption from the host lying significantly above that relation.  
The authors concluded that Na I D measurements were useful only to determine low dust extinction when the absorption was not detectable.
In addition, temporal evolution of these lines could signify changes in ionization balance induced by the SN radiation field as shown for Na I D \citep{2007Sci...317..924P,2009ApJ...693..207B}, and it has been attributed as an evidence for the presence of CSM.
Moreover \cite{2011Sci...333..856S,2014MNRAS.443.1849S}, studying the shift of the absorption line minima from their expected position in the rest-frame, found that a significant fraction of SNe Ia show blueshifted structures, suggesting gas outflows from the SN progenitor systems.
The blueshift seems to be related to the strength of the Na I D absorption, as well as with the color of the SN \citep{2013MNRAS.436..222M}.

\begin{figure}
\centering
\includegraphics[trim=1.8cm 0cm 1cm 1.4cm, clip=true,width=\columnwidth]{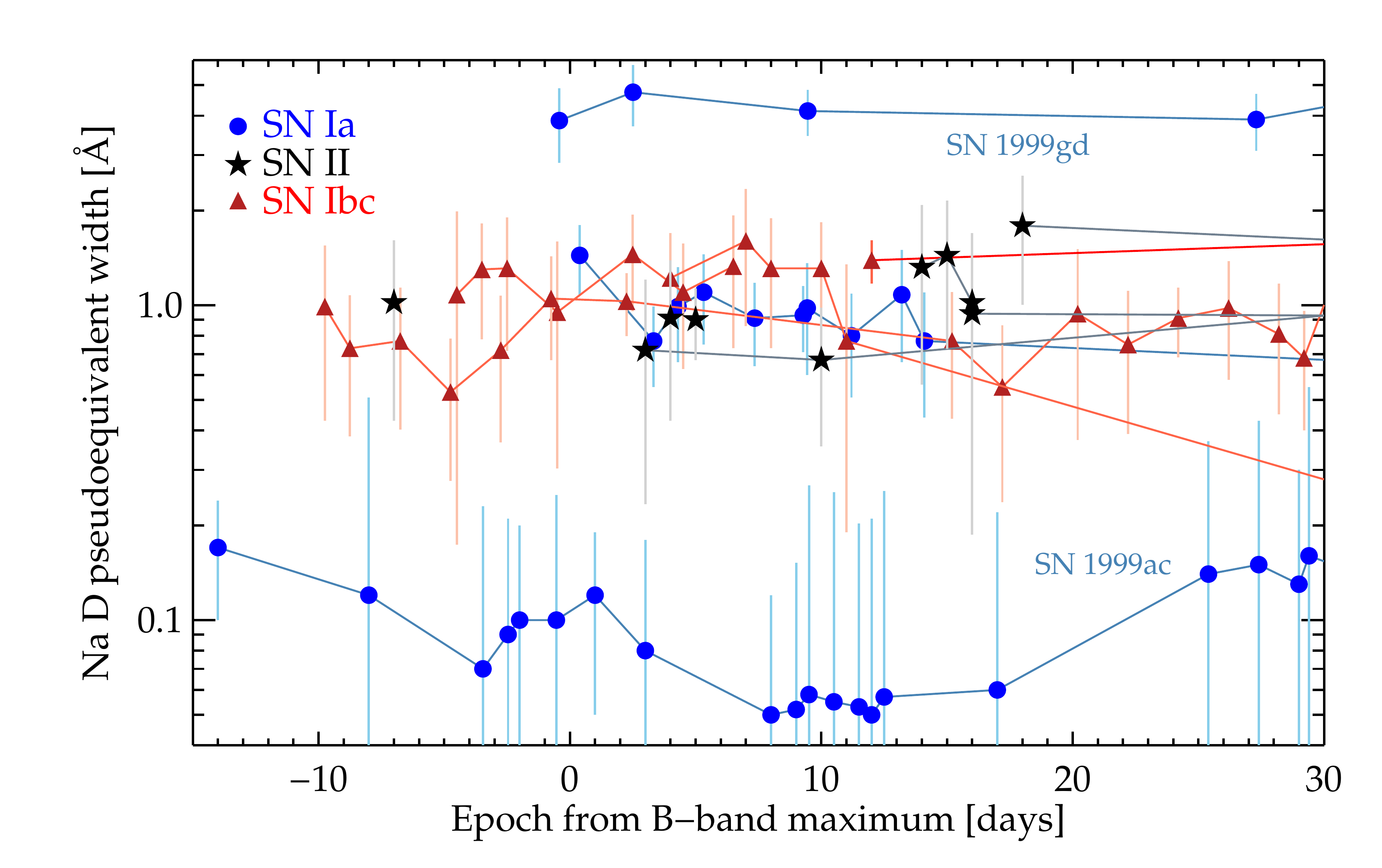}
\caption{Evolution of the pseudo-equivalent width of the Na I D absorption measured in SN spectra. Epochs are referenced to the epoch of B-band maximum light, and different colors and symbols represent different SN types.}
\label{fig:nad}
\end{figure}

In Figure \ref{fig:nad} we show our measurements of the Na I D pWs, following the procedure described in \cite{2016MNRAS.457..525G}, in all spectral epochs for the 13 SNe with spectroscopy available.
With the exception of two SNe Ia, 1999gd and 1999ac, all other SNe show relatively constant values around 1 \AA, with the lowest scatter occurring at maximum and during the following two weeks.
No clear evolution of Na I D pW with time is seen in the Figure. 

Although SN 1999gd is the object showing the highest Na I D values, the \av~from the Balmer decrement at its location was consistent with zero, and we were only able to put an upper limit of CO at its position. 
This SN has been shown to suffer a significant amount of extinction (e.g. \citealt{2015ApJ...806..134M} found A$_V\sim$ 1.280 mag from comparisons to template spectra), and has been also previously typed as a high-velocity gradient SN Ia which are also intrinsically redder \citep{2014ApJ...795..142G}.
The fact that from SN spectroscopic and photometric observations we see this object to have suffered extinction, while from the local host galaxy optical and millimeter spectra we find no indications of extinction, may suggest that the source of reddening is either some material that was around the progenitor star before the explosion, or nearby interstellar dust that is first ionized and then destroyed by the SN shock.
The high Na I D pW would support this scenario of SN Ia 1999gd being extincted by nearby dust from the ISM or CSM (more indications in that direction in section \ref{sec:phot}).
On the other hand, SN 1999ac is the object showing the lowest Na I D pW values. In this case non-detections of both \av~and CO were measured at its location, and it was classified as a core normal SN Ia.
According to our results and following \cite{2013ApJ...779...38P} conclusions, we could argue that this SN suffered no significant extinction from the ISM and shows no signs of nearby ISM/CSM.
Only SN Ia 2001en had clear local CO detection. It shows no special behaviour in Na I D with respect to other SNe, which suggests the ISM is responsible for its reddening.
SN 1989A only has one spectrum available at +88d after max and the Na I D evolution cannot be studied.
Regarding SNe II, half of the sample has only 1 spectrum and although we cannot study the evolution of the Na I D absorption, we have also put these single measurements all around 1 \AA~in Figure \ref{fig:nad}.
SN 2005ip is a SN IIn, so it is supposed to have strong interaction with CSM. Although it showed the higher local \av~value and CO was detected at its position, no specially stronger Na I D absorption are detected in its spectra compared to other objects ($\sim 1 \AA$).
SN 2005en Na I D pW follows the average evolution, and SN 2005ci has the higher Na I D absorption, but lower \av~than other SNe II, and local CO upper limit.
Although all SNe Ibc have higher \av~and \nh~values than other SN types, they do not show significantly higher Na D pW compared to other SN types.

\subsection{Photometric method: SN colors} \label{sec:phot}

The nature of red colors towards SNe is still debated. It is not clear what is intrinsic to the SN and what is due to reddening from material in the line of sight. 
In this section we explore a photometric method to estimate the reddening towards SNe, as well as any other evidence for CSM, from a photometric perspective.

\subsubsection{SN Ia color parameter}

Particularly for SNe Ia, an estimation of the extinction can be obtained from a template fit to the multicolor light curve. 
Some SN Ia light curve fitters (e.g MLCS, \citealt{2007ApJ...659..122J}) account explicitly for the visual extinction when standardizing SN Ia peak brightness. On the other hand, fitters like SiFTO \citep{2008ApJ...681..482C} adjust SN Ia maximum brightness in each filter from which an observed color parameter is obtained that includes both extinction and intrinsic color.
This intrinsic color can be measured from the B-V color curve at the epoch of maximum brightness in spectral templates, and it turns out to be on average -0.016 mag for {\it normal} SNe Ia (from \citealt{2007ApJ...663.1187H} template), and 0.043 mag (from \citealt{2015MNRAS.448.2766B} templates) for high-velocity gradient SNe Ia.

\begin{figure}
\centering
\includegraphics[trim=0.6cm 0cm 0.3cm 0.6cm, clip=true,width=\columnwidth]{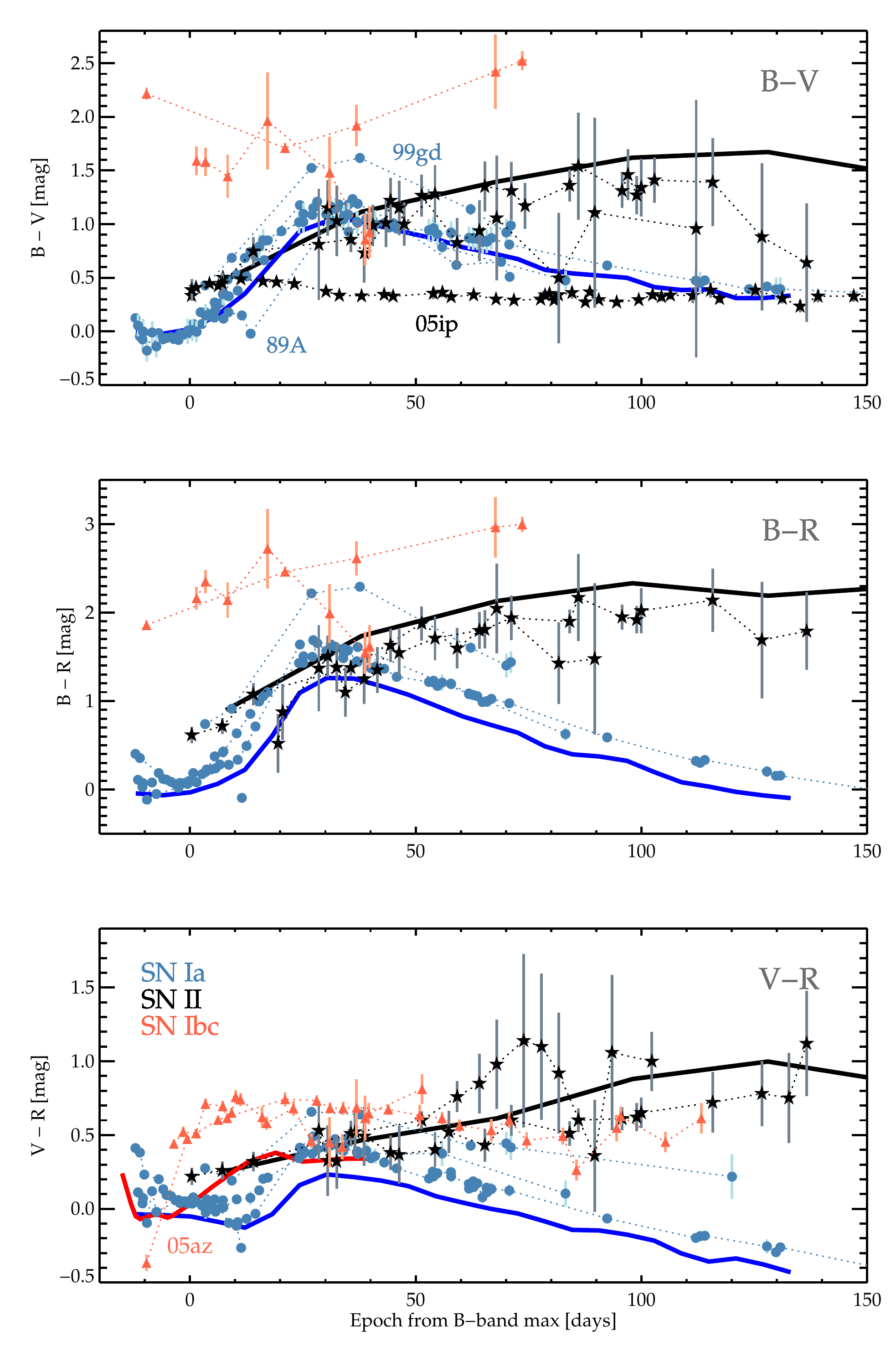}
\caption{B-V, B-R, and V-R color curves for SNe in our literature sample. Overplotted are the average color curves of \protect\cite{2013ApJ...772...19F} for SNe Ia, \protect\cite{2016AJ....151...33G} for SNe II, and \protect\cite{2011ApJ...741...97D} for SNe Ibc.}
\label{fig:color}
\end{figure}

For four SNe Ia in our sample (1989A, 1999ac, 1999gd, and 2001en) we can fit their publicly available light curves using SiFTO and obtain their color parameters. 
Taking out the contribution due to the intrinsic variation, which is assumed to be of high-velocity kind for SN 1999gd and normal for the other three SNe, we are left with an estimation of the host galaxy reddening. 
As in the previous section, the highest color excess correspond to SN 1999gd. 
It is above the typical values of -0.2 $<$ E(B-V) $<$ 0.2 (e.g. \citealt{2014ApJ...795..142G}), so that the host-galaxy extinction appears to be significant from a photometric point of view. 
Again, since no extinction was measured from the ISM, nearby material, ISM or CSM, may be responsible of causing this reddening.
The opposite case happens for SN 1989A, where we measured \av=0.27 mag, and upper limit on \nh$<$=0.79~10$^{-21}$ cm$^{-2}$, but only 0.05$\pm$0.02 color excess. This is one of the SNe in our sample that happened closer to the center of its host galaxy (5.9 kpc).
The SN location within the galaxy has been shown to be related to the amount of extinction estimated from light-curve fitting \citep{2012ApJ...755..125G}.
A possible scenario would be that this SN resulted from a foreground star with respect the galaxy disc, hence it is not significantly affected by extinction.
However, our environment measurements take into account the whole column at that position.
This is also the case for SN 1969C (2.8 kpc), for which we were able to measure \av~(at all other SN Ia locations there was not gas and we could only out upper limits). 
For SN 2001en, from its color curve it does not seem to suffer extinction and this agrees with the measurement of local \av~from Balmer decrement that was consistent with zero, but at its location we found the highest CO detection in the SN Ia sample. This is reinforced by the fact that it exploded far from the galaxy center (9.27 kpc), in a location not surrounded by warm gas but with CO present.
The low color excess for SN 1999ac is in agreement with zero \av~and CO non-detection, and also by the fact that it explode far in the outskirts of its host (11.87 kpc).

\subsubsection{Supernova color curves}

Multi-band light curves allow the study of the color characteristics and its temporal evolution, which mainly represent the temperature evolution of the SN, but also other physical parameters such as its ionization state evolution. 
It has already been shown that scatter in intrinsic color evolution exists, but keeping this caveat in mind, and assuming that those objects with the bluest colors suffer little to no reddening, one can interpret the color excess as an indication of the amount of host galaxy extinction.

\begin{figure*}
\centering
\includegraphics[trim=0.cm 0cm 0.cm 0.cm, clip=true,width=\textwidth]{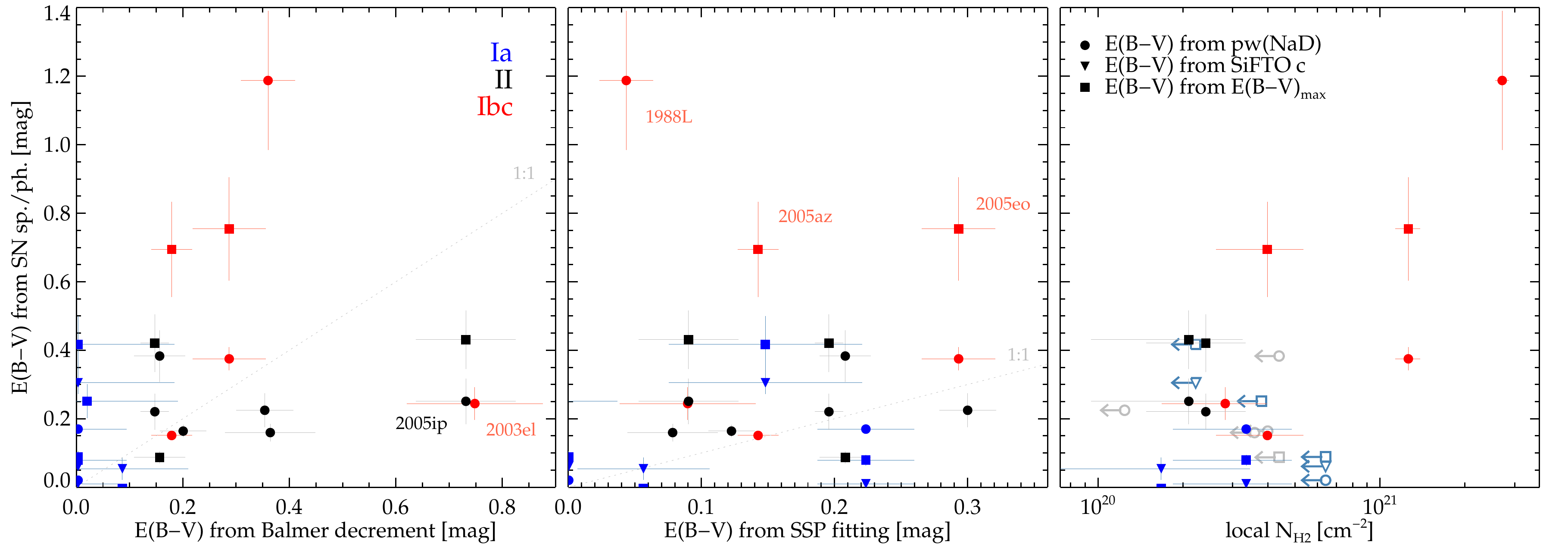}
\caption{Relation between B-V color excess measurements from SN photometry and spectra with respect different ISM reddening measurements from Balmer line decrement (left) and SSP fitting to the continuum (mid) in CALIFA optical spectra, and \nh~from EDGE maps (right).
When relations are similar, this can be understood as a SN environment clean of nearby dust, otherwise, some other factor is in play such as CSM interaction or some very peculiar intrinsic color.}
\label{fig:avs}
\end{figure*}
In Figure \ref{fig:color}, we construct B-V, B-R, and V-R color curves for the literature SN sample that were observed with these bands, with respect to the epoch of B-band light curve maximum.
On top of those, we plot average MW corrected SNe Ia \citep{2013ApJ...772...19F}, SNe II \citep{2016AJ....151...33G}, and SNe Ibc \citep{2011ApJ...741...97D} color curves.
We note that for SNe Ibc only the V-R color curve is available, and it is the only one that has also been corrected for host galaxy extinction, thus being the one better representing a non-extinction affected color curve.
To solve this issue for SNe Ia, we shifted each of the three color curves to have at maximum the same color as the \cite{2007ApJ...663.1187H} template in that color.
For SNe II, we selected the color at the epoch of B-band maximum brightness of the SN II with the bluest color, or the bluer surrounding of all the color curves in \cite{2016AJ....151...33G} sample.
In all cases, we find a few objects further than 2 sigma from the average color curve. These objects probably suffer important levels of extinction. 
On the other hand, objects located closer to the averaged curves can be interpreted as having little host-galaxy reddening. 

Typically the color evolution for SNe Ia increases steadily after maximum due to the drop in temperature, which shifts the peak of the spectral energy distribution to redder wavelengths.
At around 30 days post-maximum the color curve turns back to bluer colors, showing a constant decrease known as the {\it Lira law regime} \citep{1996Lira}.
At these epochs, all SNe Ia seem to have a narrow color range with a constant decline rate although with some variations. Fast decliners also show higher Na I D pW, have redder colors, and prefer lower R$_V$ reddening laws, indicating possible relation to CSM \citep{2013ApJ...772...19F}.
We clearly see that SN 1999gd is significantly above the average color curve by a half or a full magnitude, and also show a faster decline than other SNe Ia.
SN 2001en and SN 1969C (only in B-V) are also slightly above the average.
SN 1999ac follows the average Ia color curve, but compared to SN 2001en, it is clearly below and closer to the average color curve. 
Its color curve is flatter and it also has lower Na I D values than 2001en.
SN 1989A is following the average behaviour although a bit below.
All these results for SNe Ia are consistent with those from the LC fits.

For SNe II, the initial increase in the first weeks is followed by a less pronounced and then almost flat evolution because the temperature conditions at the photosphere remain similar due to the recombination of H happening during this phase.
After $\sim$100 days post-maximum, when the radioactive phase starts, the color curves become flatter because in this phase the SN II photometric evolution, which depends on the $^{56}$Co decay, is approximately the same in all bands.
SN 2005au follows the average color curves until 50 days after maximum, when it shows a strange behaviour although it has large errors.
SN 2005ci follows the average but always a bit below the curve.
Both objects have local \av$\sim$0.5 mag, but only SN 2005au had CO detection.
SN IIn 2005ip has a flat color curve B-V around 0.4 mag. Similar to the SN IIn 2003bj in \cite{2016AJ....151...33G}, it does not follow the standard SN II color curve.
It has the highest local \av~value, and also CO was detected at its position. 

\cite{2011ApJ...741...97D} concluded that SNe Ibc had more evidences for significant host galaxy extinction than SNe II. They use this argument to suggest that SNe Ibc are more embedded in the star-forming regions than SNe IIP, which is consistent with environmental results \citep{2012MNRAS.424.1372A,2014A&A...572A..38G}.
SN 2005eo is far above ($\sim$0.5 mag) the non-extinguished average color curve.
SN 2005az V-R pre-maximum is bluer than the average, but later becomes 0.5 mag redder than the unextinguished average SN Ibc color curve.
We do not have non-extinguished B-R and B-V color curves, but these two supernova have color curves that are larger than up to 2 mag than SNe of other types.
Both SNe have non-zero \av, CO detection, and constant Na I D pW of $\sim$1\AA.

\subsection{Method comparisons}

Here we compare the B-V color excess obtained through different estimations from SN data, to the E(B-V) obtained from environmental optical data through the Balmer decrement and SSP fitting, and from molecular cold gas at SN locations. 

Taking the median value from our Na I D pW estimations for each SN, we calculated the observed color excess, E(B-V), using the \cite{2012MNRAS.426.1465P} relation.
On the other hand, we measured the E(B-V) from the photometric colors as the difference between the average (B-V) color and the SN color at the epoch of B-band maximum.
When possible, we performed this measurement independently for the 3 colors available (B-V, V-R, and B-R), converted to E(B-V) using a \cite{1999PASP..111...63F} extinction law,  and then averaged out.
For the four SNe Ia with LC available, we also included E(B-V) estimated from the color parameter.
In Figure \ref{fig:avs}, we show the comparisons between the E(B-V) obtained from SN spectroscopic and photometric methods to E(B-V) estimated in the environment.
It can be clearly seen that E(B-V) from Na I D pW does not have a 1-to-1 relation to the E(B-V) and N(H2) measured in galaxy spectra at the SN position. 
SN 1988L (Ib) is clearly offset towards very high extinction values measured from Na I D absorption, while SN 2003el (Ic) and SN 2005ip (IIn) have significant lower E(B-V) values as proxied from Na I D pW than E(B-V) from the Balmer decrement.
Na I D pW seems to better follow the E(B-V) measurement from the SSP fitting, with the only exception of SN 1988L (Ib).
Given the large amount of CO non-detections it is difficult to establish a clear relation, but those 3 SNe Ibc that occurred at positions with the largest \nh~showed also large color excess. 
SN 1988L showed both the highest E(B-V) and \nh~at its location.

When looking at the E(B-V) estimated from SN colors, we find similar disagreement when compared to E(B-V) from Balmer decrement.
We estimated significantly higher E(B-V) at SN 2005az (Ic) and SN 2005eo (Ic) locations from light-curve colors than from Na I D pW relation.
Thus, this results would favor the picture where the reddening of these objects is produced by dust that is local to the SN.
These two SNe are offset when comparing with the environmental E(B-V) from both the gas and the stellar component, while they both occurred at high \nh~locations.

Althouhg SN 1999gd (Ia) was the object with the largest Na I D pW, and also the redder SN Ia, we detected no extinction from the environment. Therefore, it is also far from the 1-to-1 relation in the Figure. Some other effect has to be responsible of reddening the light from this SN, specially when all other SNe Ia have no differences as pronounced as 1999gd. 

In summary, Figure \ref{fig:avs} clearly shows that extragalactic extinction measurements are highly uncertain. Although some trends can be found in the data, given different extinction mechanisms, it is difficult to arrive to strong conclusions, pointing that extinction toward extragalactic objects is not yet well solved.
However, in this work we have found that extinction measured in the optical from SN data is usually higher than measured in the environment after the SN faded, which indicates that the amount of dust was higher before/during the SN explosion. 
These differences are probably caused because dust is destroyed to some extent by the SN shock (when it was located nearby the progenitor star), although resolution effects may be introducing some dilution effect in the measurement from the environment.

\section{Conclusions} \label{sec:conc}

We presented a study of 23 galaxies that hosted 26 SNe (7 SNe Ia, 12 SNe II and 7 SNe Ib/c) which had spatially resolved observations in the millimeter by the EDGE survey and in the optical by the CALIFA survey.
We focused on the integrated and local properties of host galaxies at the position of each SN with the aim to find possible correlations between the SN positions in the galaxies and their type, and compare these results with similar observations in the optical. Below we list our conclusions:

\begin{itemize}
\item SNe Ibc tend to explode at locations with larger amounts of molecular gas compared to both SNe II and SNe Ia, which show similar distributions. This indicates that SNe Ibc occur at places with large cold gas reservoir where star formation is ongoing.

\item The difference in the average values of the SN Ibc and SN II distributions, with the current numbers, is even larger for the local S$_{CO}$ (gas reservoir, future star formation) than for the local H$\alpha$ flux (current star formation), where they are statistically indistinguishable. This reinforces the fact that SNe Ibc are more associated with SF-environments, and their progenitor stars have shorter lifetimes and thus larger masses.

\item We find a correlation between the total H$_2$ column density and the total optical extinction of 2.10 $\pm$ 0.47 cm$^{-2}$ mag, which is consistent to previous reports in the literature.

\item SNe Ibc have both larger H$_2$ column densities and dust reddening at their locations than other SN types. While SNe Ia tend to occur at places with lower \nh~and \av, SNe II occur at places with low \nh~but larger amount of dust.

\item When comparing environmental proxies for extinction to estimations extracted from SN photometry and spectroscopy, such as Na I D absorption pW, SN Ia color parameter, and SN color curves, we do not find a significant relation, which can be interpreted as an evidence for some other effect in play. When the extinction measured directly from SN observations is higher than the estimation from the environment when the SN faded, as we find for some objects, it can be an evidence of CSM interaction.

\item SNe Ibc in our sample are also the SN type showing the reddest colors, and for this SN type the offset between the 1-to-1 relation between the extinction estimated from SN and environmental data is larger, so we conclude that they may be the SN type being more affected by CSM interaction. Given that the progenitors of these SNe are expected to lose their outer envelopes before explosion, our finding would be expected.
Although SNe II show intermediate behaviour, SNe Ia and SNe Ibc seem to behave in a clear different manner. 

\item While SN Ia cluster fall above the proportional line to the E(B-V) measured via the Balmer decrement, E(B-V) measured via SSP fitting (sensitive to stellar continuum) is relatively consistent with E(B-V) from SN Ia properties.

\item SN Ia 1999gd is the redder SN Ia and have the higher Na I D pW in our sample. With our results, and following previous works, we also conclude that this object was affected by CSM extinction, given that their environment has no signs of ISM extinction.

\end{itemize}

This work is a benchmark for future studies using the capabilities of much better resolution optical IFS, (such as MUSE at the VLT, \citealt{2016MNRAS.455.4087G}) and by millimeter observatories with larger light collecting areas such as the Atacama Large (Sub)Millimeter Array (ALMA).
Deeper \CO~observations or in other optically thin tracers such as $^{13}$CO and $^{18}$CO with ALMA will be crucial to establish if any of the above scenarios still stands, and larger samples are needed to arrive to strong conclusions on the molecular gas properties at SN locations.

\section*{Acknowledgments}

L.G. was supported in part by the US National Science Foundation under Grant AST-1311862. 
Support for LM and SGG is provided by the Ministry of Economy, Development, and Tourism's Millennium Science Initiative through grant IC120009, awarded to The Millennium Institute of Astrophysics, MAS.
H.D. acknowledges financial support from the Spanish Ministry of Economy and Competitiveness (MINECO) under the 2014 Ram\'on y Cajal program MINECO RYC-2014-15686.
The work by KM has been supported by JSPS KAKENHI Grant 26800100 and by JSPS Open Partnership Bilateral Joint Research Project between Japan and Chile.
S.P acknowledges support by FONDECYT grant 3140601 and the Millennium Nucleus on protoplanetary disks RC130007 (Chilean Ministry of Economy).
SFS thanks the CONACYT-125180, DGAPA-IA100815 and DGAPA-IA101217 projects for providing him support in this study.
This research has made use of the CfA Supernova Archive, which is funded in part by the National Science Foundation through grant AST 0907903.

\bibliographystyle{mnras}
\bibliography{edge}

\end{document}